\documentclass[twocolumn]{aastex7}

\usepackage{tablefootnote}
\usepackage{multirow}
\usepackage{makecell, mathtools}
\usepackage{hyperref}
\usepackage{subfigure}
\usepackage{enumitem}

\submitjournal{ApJ}

\shorttitle{AT\,2022cmc in context}
\shortauthors{Hammerstein et al.}

\begin{document}

\title{The Jetted Tidal Disruption Event AT\,2022cmc: Investigating Connections to the Optical Tidal Disruption Event Population and Spectral Subclasses Through Late-Time Follow-up}

\author[0000-0002-5698-8703]{Erica Hammerstein}
\email[show]{ekhammer@berkeley.edu}
\affil{Department of Astronomy, University of California, Berkeley, CA 94720, USA}

\author{S.~Bradley Cenko}
\email{brad.cenko@nasa.gov}
\affil{Astrophysics Science Division, NASA Goddard Space Flight Center, Mail Code 661, Greenbelt, MD 20771, USA}

\author[0000-0002-8977-1498]{Igor Andreoni}
\email{igor.andreoni@unc.edu}
\affil{University of North Carolina at Chapel Hill, 120 E. Cameron Avenue, Chapel Hill, NC 27514, USA}

\author[0000-0002-0326-6715]{Panos Charalampopoulos}
\email{pachar@utu.fi}
\affil{Department of Physics and Astronomy, FI-20014, University of Turku, Finland}

\author[0000-0002-7706-5668]{Ryan Chornock}
\email{chornock@berkeley.edu}
\affil{Department of Astronomy, University of California, Berkeley, CA 94720, USA}

\author[0000-0003-4768-7586]{Raffaella Margutti}
\email{rmargutti@berkeley.edu}
\affil{Department of Astronomy, University of California, Berkeley, CA 94720, USA}
\affil{Department of Physics, University of California, 366 Physics North MC 7300, Berkeley, CA 94720, USA}

\author[0000-0002-9700-0036]{Brendan O'Connor}
\email{boconno2@andrew.cmu.edu}
\affil{McWilliams Center for Cosmology and Astrophysics, Department of Physics, Carnegie Mellon University, Pittsburgh, PA 15213, USA}

\author[0000-0001-6797-1889]{Steve Schulze}
\email{steve.schulze@northwestern.edu}
\affiliation{Center for Interdisciplinary Exploration and Research in Astrophysics (CIERA), Northwestern University, 1800 Sherman Avenue, Evanston, IL 60201, USA}

\author[0000-0003-1546-6615]{Jesper Sollerman}
\email{jesper@astro.su.se}
\affil{Department of Astronomy, The Oskar Klein Center, Stockholm University, AlbaNova University Center, SE 106 91 Stockholm, Sweden}

\author[0000-0002-3927-5402]{Sudhanshu Barway}
\email{sudhanshu.barway@iiap.res.in}
\affil{Indian Institute of Astrophysics, II Block Koramangala, Bengaluru 560034, India}

\author[0000-0002-6112-7609]{Varun Bhalerao}
\email{varunb@iitb.ac.in}
\affil{Department of Physics, Indian Institute of Technology Bombay, Powai 400076, India}

\author[0000-0003-3533-7183]{G.~C.~Anupama}
\email{gca@iiap.res.in}
\affil{Indian Institute of Astrophysics, II Block Koramangala, Bengaluru 560034, India}

\author[0000-0003-0871-4641]{Harsh Kumar}
\email{harshkumar@fas.harvard.edu}
\affil{Center for Astrophysics \textbar\,Harvard \& Smithsonian, 60 Garden Street, Cambridge, MA 02138-1516, USA}

\author[0000-0002-6894-1267]{Ester Marini}
\email{ester.marini@inaf.it}
\affil{INAF -- Osservatorio Astronomico di Roma, Via Frascati 33, 00078 Monteporzio Catone, Rome, Italy}

\author[0000-0002-7409-8114]{Diego Paris}
\email{diego.paris@inaf.it}
\affil{INAF -- Osservatorio Astronomico di Roma, Via Frascati 33, 00078 Monteporzio Catone, Rome, Italy}

\author[0000-0001-8472-1996]{Daniel A.~Perley}
\email{d.a.perley@ljmu.ac.uk}
\affil{Astrophysics Research Institute, Liverpool John Moores University, IC2, Liverpool Science Park, 146 Brownlow Hill, Liverpool L3 5RF, UK}

\author[0000-0002-8860-6538]{Andrea Rossi}
\email{andrea.rossi@inaf.it}
\affil{INAF -- Osservatorio di Astrofisica e Scienza dello Spazio di Bologna, via Piero Gobetti 93/3, I-40129 Bologna, Italy}

\author[0000-0001-6747-8509]{Yuhan Yao}
\email{yuhanyao@berkeley.edu}
\affil{Miller Institute for Basic Research in Science, 468 Donner Lab, Berkeley, CA 94720, USA}
\affil{Department of Astronomy, University of California, Berkeley, CA 94720, USA}

% \correspondingauthor{Erica Hammerstein}

\begin{abstract}
AT\,2022cmc is the first on-axis jetted tidal disruption event (TDE) to be discovered at optical wavelengths. The optically bright nature of AT\,2022cmc presents an unprecedented opportunity to place this jetted TDE in the context of the larger optically selected thermal TDE population and explore potential connections to optical TDE subclasses, particularly the class of luminous TDEs that lack optical spectral features. In this work we present late-time optical observations of AT\,2022cmc, both imaging and spectroscopy, that extend the optical dataset to $\sim 160$ days from the first detection in the observed frame. The light curve clearly evolves from red to blue, which we interpret as a transition from a non-thermally dominated spectral energy distribution (SED) to thermally dominated SED. By accounting for the non-thermal emission evident in the optical SED at early times, we extract the properties of the thermal emission and compare to a sample of optically selected thermal TDEs. We find that the properties of AT\,2022cmc are consistent with previous correlations found for the evolution and properties of thermal TDEs, with the thermal properties of AT\,2022cmc aligning with the class of featureless and luminous TDEs. The confirmation of this similarity motivates the importance of prompt and multi-wavelength follow-up of featureless and luminous TDEs in order to further explore the connection they have with jetted TDEs.
\end{abstract}

\keywords{\uat{Relativistic jets}{1390} --- \uat{Black holes}{162} --- \uat{Tidal disruption}{1696} --- \uat{Transient sources}{1851}}

\section{Introduction} \label{sec:intro}
Every so often ($\sim 10^{4}$ years) in the nucleus of a galaxy, a star will wander too close to the Supermassive black hole (SMBH) lurking there and will be subsequently torn apart by the tidal forces and accreted by the SMBH \citep{Hills75, Frank76}. These tidal disruption events (TDEs) create luminous flares of radiation visible from Earth. Samples of TDEs have now been discovered across the electromagnetic spectrum from X-ray \citep[e.g.,][]{komossa99a, Sazonov2021} to optical \citep[e.g.,][]{gezari12b, Hammerstein23, Yao2023}, infrared \citep[e.g.,][]{Masterson24}, and radio \citep[e.g.,][]{Somalwar23}.

The advent of all-sky optical surveys, such as Pan-STARRS \citep{chambers16}, ASAS-SN \citep{shappee14}, and the Zwicky Transient Facility \citep[ZTF,][]{Bellm2019, Graham19}, in the last $\sim20$ years has led to the majority of current TDE discoveries being made at these wavelengths. These optically selected TDEs are typically dominated by a hot ($10^4$--$10^5$ K), thermal continuum with rise timescales of $\sim30$ days, peak absolute magnitudes of $M_r \sim -17$ to $-22$ mag \citep[e.g., see Figure 9 of][]{Yao2023}, and fade timescales of 200--400 days \citep[e.g.,][]{vanVelzen21, Hammerstein23}. Optically selected TDEs show a broad range of emission at other wavelengths, including a variety of behaviors in the X-ray \citep[e.g.,][]{vanVelzen21, Hammerstein23, Guolo23, Yao2024_22lri}. The optical spectra of TDEs are most often characterized by broad emission features \citep[e.g.,][]{Hammerstein23, Charalampopoulos22}, with \citet{vanVelzen21} formalizing a classification scheme based on the observed lines being primarily Balmer or \ion{He}{2}, known as TDE-H or TDE-He respectively, or a combination with occasional evidence for Bowen fluorescence emission, known as TDE-H+He. \citet{Hammerstein23} put forth an additional spectroscopic class of TDEs called TDE-featureless that show no discernible emission lines or spectroscopic features present in the three other classes, although host galaxy absorption lines can be observed. This featureless class of TDEs is characterized by higher peak luminosities ($M_r \sim -21$ mag), peak blackbody temperatures ($T_{\rm BB} \gtrsim 10^{4.4}$ K), and peak blackbody radii ($R_{\rm BB} \gtrsim 10^{15.4}$ cm). Their host galaxies are often more massive, potentially implying more massive black holes, and redder than typical TDE hosts, which tend to favor ``green'' galaxies \citep[e.g.,][]{LawSmith17, Hammerstein2021}.

TDEs discovered at other wavelengths, such as in the soft X-ray, are also characterized by thermal emission, though the thermal continuum in X-ray selected events is typically much hotter than their optically selected counterparts \citep[see][for a review]{Gezari2021}. A very small fraction of TDEs have been discovered through non-thermal emission from an on-axis, collimated, relativistic jet (hereafter ``jetted TDE''). Three of these objects were discovered more than a decade ago by the hard X-ray Burst Alert Telescope (BAT; \citealt{Barthelmy2005}) aboard the \textit{Neil Gehrels Swift Observatory} \citep{Gehrels2004}. These candidates include Sw J1644+57 \citep{bloom11, burrows11, levan11, zauderer11}, Sw J2058+05 \citep{Cenko2012, pasham15}, and Sw J1112-82 \citep{Brown2015, Brown2017}. Some of these jetted TDEs have shown faint or late-time optical counterparts. Jetted TDEs provide an important opportunity to study the launching of relativistic jets by SMBHs, the jet emission mechanism, and the jet composition \citep[for a review, see][]{DeColle2020}.

Recently, a fourth candidate jetted TDE was reported. In contrast to the previous three candidates, AT\,2022cmc (ZTF22aaajecp) was discovered by ZTF as an optical transient \citep{Andreoni2022}. The first detection in the optical on 2022 February 11 was shortly followed by detections in the radio \citep{Perley22}, sub-millimeter \citep{Perley22_SMM}, and X-ray \citep{Pasham22}. The emission across wavelengths is exceptionally luminous, and the redshift ($z = 1.193$) provided by follow-up spectroscopy \citep{Tanvir22} implies an absolute optical luminosity of $M_i \approx -25$ mag at peak \citep{Andreoni2022}. The follow-up observations revealed remarkable similarities to the jetted TDE Sw J1644+47, including long-lived X-ray emission with short timescale variability and a drop in flux at late-times due to jet shut-off \citep{Eftekhari24}, and corresponding radio and infrared counterparts \citep{bloom11, burrows11, levan11}. Sw J1644+57 was not detected in the optical or ultraviolet until much later \citep{Levan2016}, although this is unsurprising due to the large inferred host galaxy extinction \citep{burrows11}. Because of these similarities, and after ruling out other possible transients such as a kilonova, a luminous fast blue optical transient (LFBOT), blazar, or a $\gamma$-ray burst (GRB), the interpretation of AT\,2022cmc as a jetted TDE is favored \citep{Andreoni2022, Pasham23, Rhodes23}.

Interestingly, AT\,2022cmc's UV, optical, and IR light curve is characterized by a red color which turns bluer after several days post-peak (hereafter ``red phase'' and ``blue phase'', respectively). This contrasts with the typical evolution of optical TDEs, which show little evolution from their initial blue colors throughout their light curves \citep[e.g.,][]{vanVelzen21, Hammerstein23, Yao2023}. \citet{Andreoni2022} interpret this fast-fading red flare as synchrotron emission resulting from the jet interaction with the circumnuclear medium. The slower-evolving, blue, thermal optical/UV emission is thought to have origins similar to non-jetted TDEs, which may come from the reprocessing of X-ray emission originating in the accretion disk to the UV and optical by a wind or outflow \citep{loeb97, Guillochon2014, Dai2018}, or shocks and subsequent outflows created by intersecting stellar debris streams \citep{Piran2015, Jiang2016, lu20}. The blue component of AT\,2022cmc's optical/UV light curve indeed shows similar properties to non-jetted TDEs, with a high rest-frame UV luminosity of $\sim10^{45}$ erg s$^{-1}$ and a blackbody temperature $\sim 3 \times 10^4$ K \citep{Andreoni2022, Yao2024_22cmc}.

The follow-up optical spectra of AT\,2022cmc, both in the red and blue phases, show a featureless continuum with evidence for host stellar absorption lines but no broad features typically associated with many TDEs. Sw J2058+05 showed a similar lack of features in its optical spectrum \citep{Cenko2012, pasham15}. Because of its featureless optical spectra and high peak optical luminosity, \citet{Andreoni2022} suggested there may be a connection between jetted TDEs and the featureless class of TDEs put forth by \citet{Hammerstein23}, which show similar high optical luminosities ($M_r \sim -22$ mag). This possible connection was further bolstered by \citet{Hammerstein23_IFU}, who found that the SMBH at the center of the featureless TDE AT\,2020qhs \citep{Hammerstein23} has a mass of $\log(M_{\rm BH}/M_\odot) = 8.01 \pm 0.82$ obtained from stellar kinematics. The high black hole mass could require that the SMBH possesses a spin ($a \geq 0.16$) that would ensure that the tidal radius remains outside the event horizon radius and produces an observable TDE, however this would require the disruption of a solar-type star as a lower density star such as a giant could still be disrupted outside of the event horizon of a more massive SMBH \citep{MacLeod2012}. This may lend further support to the possible connection between jetted TDEs and featureless TDEs, if high spin is required to launch a relativistic jet \citep[e.g.,][]{Tchekhovskoy14}. \citet{Andreoni2022} proposed that TDE-featureless objects may be off-axis jetted TDEs, though current late-time time follow-up of TDE-featureless events, particularly in the radio where late-time jet emission can be constrained, is as of yet non-existent.

Despite extensive ground-based follow-up observations, the host galaxy of AT\,2022cmc has yet to be detected. Limits on the host magnitude were obtained from forced photometry to archival imaging in $u$- and $r$-bands obtained with the MegaPrime camera on the 3.58-m Canada-France-Hawaii Telescope in 2015 and 2016. These limits are $m_u >24.19$ mag and $m_r>24.54$ mag. \citet{Andreoni2022} estimated the host properties by modeling the SED. The limits on the galaxy luminosity yielded limits on the host galaxy stellar mass of $\log (M/M_\odot) < 11.2$, which we use as the upper limit in this work. Using these limits on the host, \citet{Andreoni2022} put a limit on the SMBH mass of $\log (M/M_{\rm BH}) < 4.7\times 10^8 M_\odot$. \citet{Eftekhari24} favor lower black hole masses $< 10^5 M_\odot$ inferred from the observed X-ray luminosity (assumed to be at the transition between super-Eddington and sub-Eddington) and jet shut-off timescale instead of host galaxy limits. Deep, late-time observations of AT\,2022cmc are needed to constrain the host galaxy properties and the properties of the SMBH more strongly.

The discovery of AT\,2022cmc, a jetted TDE with a bright optical counterpart, presents an unprecedented opportunity to place this rare class of TDEs in the context of the non-jetted optical TDE population and study the potential connection between TDE-featureless objects and jetted TDEs. We therefore present additional observations of AT\,2022cmc at late times to better constrain the thermal component, as well as an analysis of the optical/UV light curve and optical spectra of AT\,2022cmc to place it within the context of other optically selected TDEs. The paper is organized as follows. We describe the observations and data reduction in Section \ref{sec:obs}. In Section \ref{sec:analysis}, we describe the optical spectra fitting methods and light curve fitting methods we employ. We present the results of both of these in Section \ref{sec:results} and end with a discussion and conclusions in Section \ref{sec:discussion}. Throughout this paper, we adopt a flat cosmology with $\Omega_\Lambda = 0.7$ and $H_0 = 70 ~\mathrm{km~s^{-1}~Mpc^{-1}}$. All magnitudes are reported in the AB system.

\section{Data \& Observations} \label{sec:obs}
\subsection{Photometry}
\begin{figure*}
    \centering
    \includegraphics[width=\textwidth]{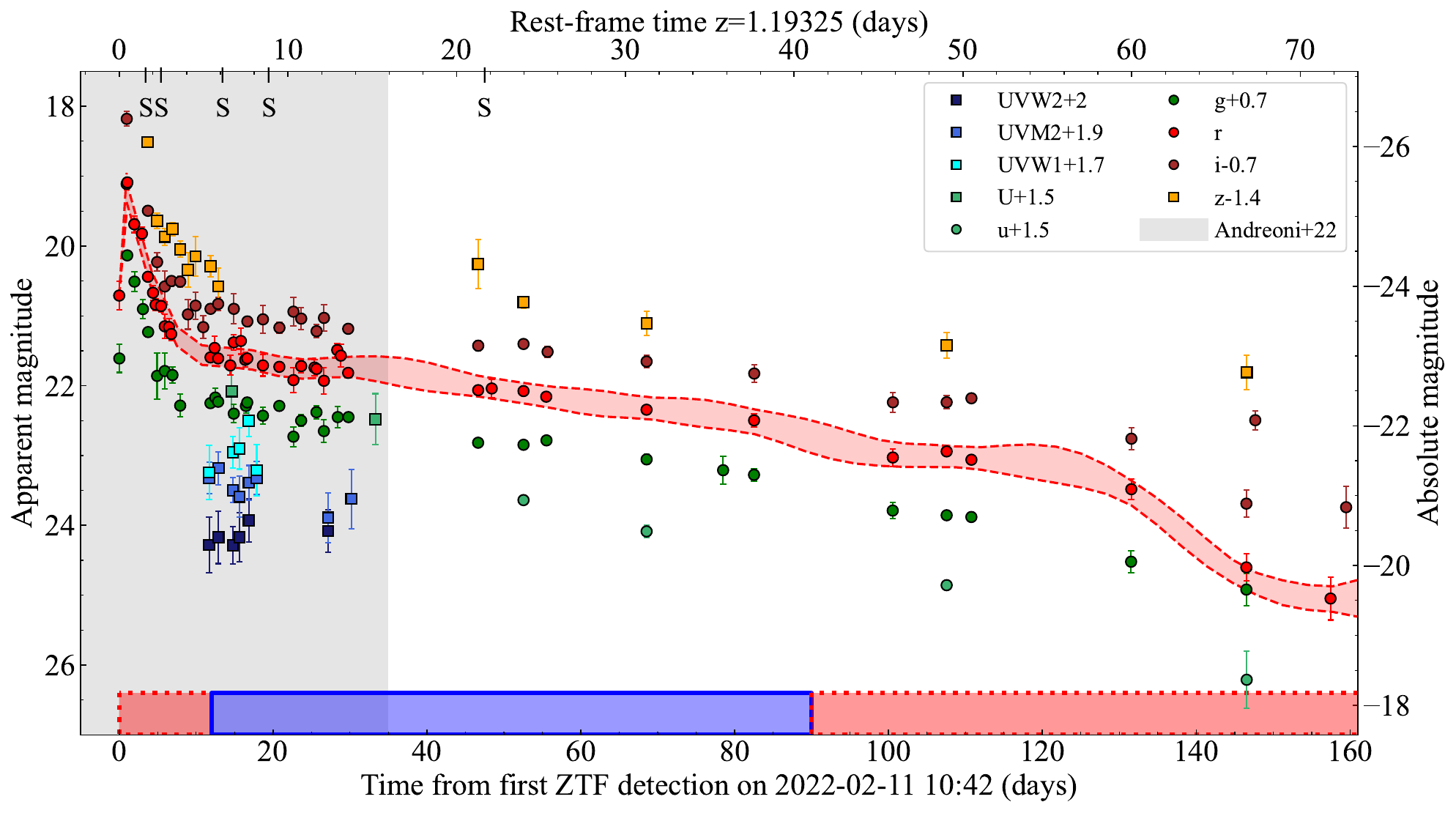}
    \caption{UV and optical light curve of AT\,2022cmc in the observed-frame. The original discovery light curve published in \citet{Andreoni2022} covers the first $\sim$35 days since discovery (observed frame) and is indicated by the gray shading. Here we present additional observations that extend the light curve up to $\sim$160 days since first detection. We show the 1-$\sigma$ band for a Gaussian process regression for the $r$-band data. Magnitudes are not corrected for Galactic extinction and are not host-subtracted as the host is not yet detected in ground-based imaging. We expect the host contribution to be minimal for the majority of the light curve. Epochs where optical spectra were taken are marked with an ``S''. On the bottom axis, we note the approximate times of the ``red'' (dotted border) and ``blue'' (solid border) phases of the light curve, according to the $g-r$ light curve in Figure \ref{fig:g-r}, with the late-time light curve becoming red once again.}
    \label{fig:lightcurve}
\end{figure*}

\begin{figure}
    \centering
    \includegraphics[width=\columnwidth]{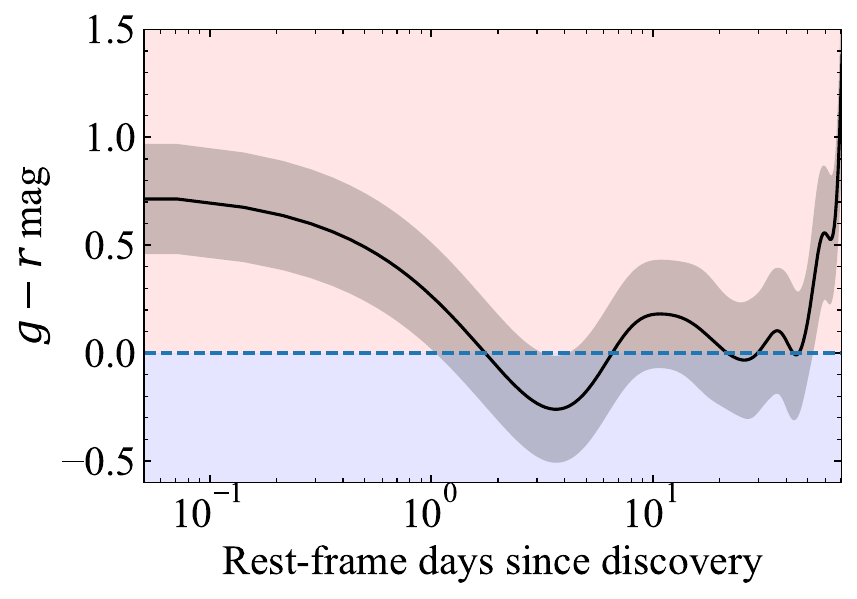}
    \caption{The $g-r$ light curve estimated from Gaussian process regression for both the $g$- and $r$-bands. The shaded region represents the 1-$\sigma$ band and we mark $g-r=0$ with the dashed line. The light curve evolves from red to blue (i.e., $g-r\lesssim0$) rapidly and remains fairly blue until $\sim50$ days, after which the light curve significantly reddens. This may be due to a larger host contribution relative to the fading thermal TDE emission at later times. Considering that \citet{Andreoni2022} constrained the host emission to $m_r > 24.54$ mag and the $r$-band light curve drops below this limit around 60 days in the rest-frame, this may be plausible.}
    \label{fig:g-r}
\end{figure}

Here we present new observations of AT\,2022cmc taken since its discovery presented in \citet{Andreoni2022}. In Figure \ref{fig:lightcurve}, we show the UV/optical light curve including the observations presented in \citet{Andreoni2022} which cover up to $\sim 35$ days from the first detection (observed frame). The light curve is characterized by a red color at early times that dominates until $\sim12$ days (observed-frame) and evolves towards a blue ($g-r\lesssim 0$ mag) color (Figure \ref{fig:g-r}). This color persists on the relatively uniform decline until $\sim60$ days where the light curve decays more rapidly.

We point the reader to \citet[][their Section 12]{Andreoni2022} for details on the reduction of the discovery light curve data. The additional data that we present here extend the light curve to $\sim160$ days from the first detection (observed frame) and include observations from the GROWTH India Telescope~\citep[GIT;][]{2022AJ....164...90K}, the Nordic Optical Telescope (NOT) Alhambra Faint Object Spectrograph and Camera (ALFOSC), Lowell Discovery Telescope (LDT) Large Monolithic Imager (LMI), Large Binocular Telescope \citep[LBT,][]{LBT} Large Binocular Camera \citep[LBC,][]{LBC}, and the Liverpool Telescope (LT) IO:O camera \citep{LiverpoolTelescope}. Standard photometric reduction procedures such as bias subtraction, flat-fielding, and cosmic ray rejection were performed on all data.

The GIT photometry was reduced using standard imaging data reduction procedures, including bias correction, flat-fielding, and cosmic ray removal with the \texttt{Astro-SCRAPPY} package. Any stacking was performed with \texttt{SWarp} \citep{bertin02, Bertin11}. Instrumental magnitudes were extracted using PSF photometry and calibrated against the PanSTARRS DR1 catalog, described in~\cite {GITpipeline} in detail.

NOT+ALFOSC (Program IDs: 64-501, 64-011, and 65-030 with PIs Sollerman and Charalampopoulos) $gri$ observations were reduced with \footnote{https://github.com/jkrogager/PyNOT}{\texttt{PyNOT}} using standard routines for imaging data. We used aperture photometry to measure the brightness of the transient which we then calibrated against the brightness of several stars from a cross-matched SDSS catalog.

We reduced the LDT+LMI data using standard data reduction techniques, including bias subtraction, flat-fielding, and cosmic ray rejection. Aperture photometry was computed using \texttt{SExtractor} \citep{Bertin96} and calibrated to the PanSTARRS catalog.

LBT+LBC imaging data were reduced using the data reduction pipeline developed at INAF - Osservatorio Astronomico di Roma \citep{Fontana2014a} which includes bias subtraction and flat-fielding, bad pixel and cosmic ray masking, astrometric calibration, and coaddition. The astrometry was calibrated against field stars in the GAIA DR3 catalog. PSF photometry was performed using \texttt{DAOPHOT} and \texttt{APPHOT} under \texttt{PyRAF/IRAF} and calibrated against the SDSS catalog.

We performed astrometric alignment and stacking of the LT IO:O observations, with exposures showing major tracking errors or poor transparency due to cloud cover being discarded. We extracted the photometry of the transient using a custom IDL routine using seeing-matched aperture photometry fixed at the transient location which was then calibrated to a set of SDSS secondary standard stars in the field.

\subsection{Spectroscopy}
\begin{figure*}
    \centering
    \includegraphics[width=\textwidth]{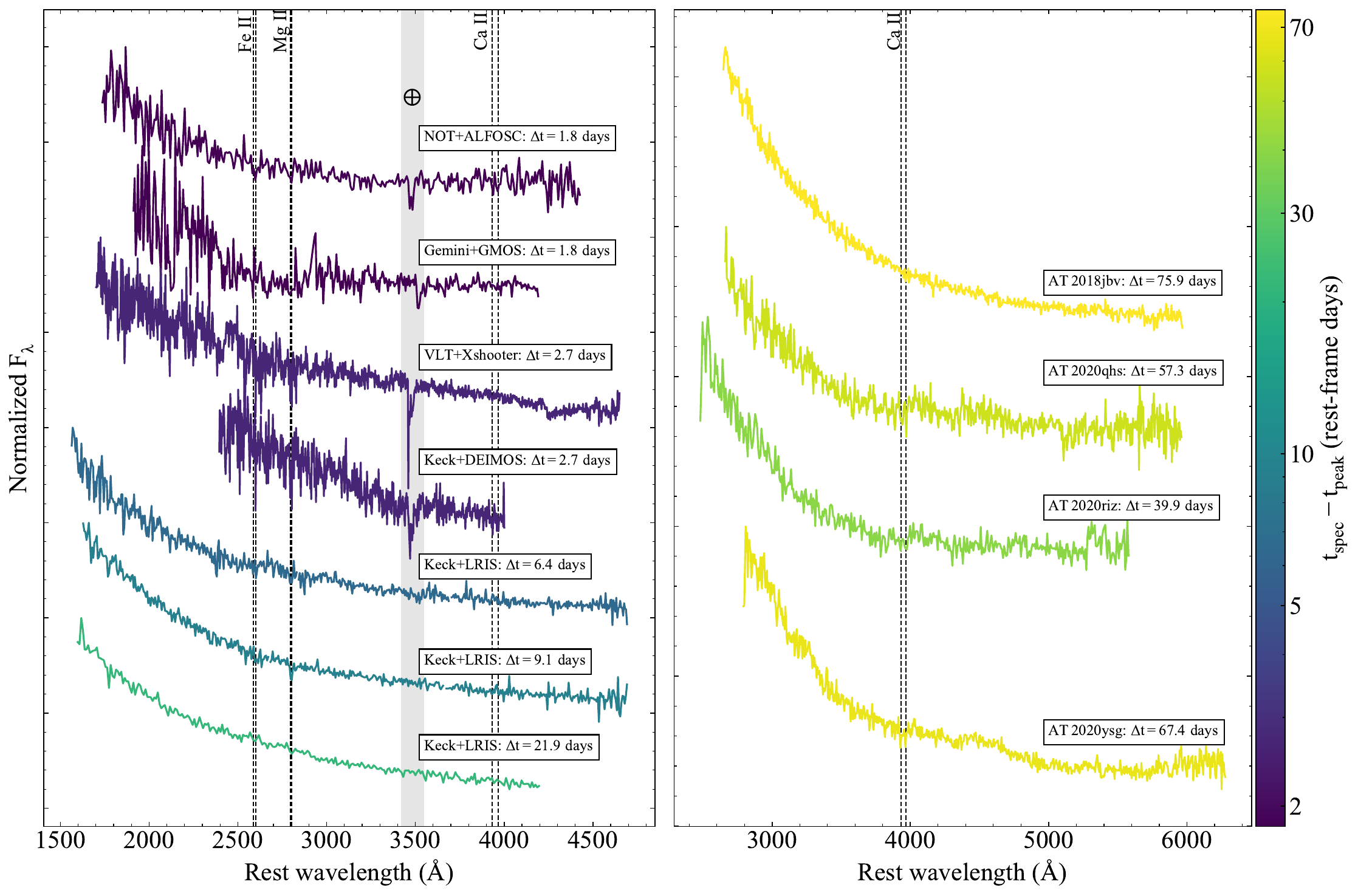}
    \caption{\textit{Left}: Optical spectra of AT\,2022cmc from \citet{Andreoni2022} with an additional spectrum from Keck/LRIS at $\Delta t=21.9$ days (rest-frame), presented here for the first time. The spectra cover both the red and blue phases of the light curve, with the red component dominating until $\sim 5$ days rest-frame, but remain featureless despite the clear evolution in the continuum. The absorption line near 3500 \AA~in the AT\,2022cmc spectra is telluric (non-astrophysical). \textit{Right}: Optical spectra of the four TDE-featureless objects in \citet{Hammerstein23}, which show remarkable similarities to the later spectra of AT\,2022cmc. The \ion{Ca}{2} host stellar absorption feature can be seen. The color of each spectrum corresponds to the approximate phase it was taken relative to peak (or first detection for AT\,2022cmc). All spectra have been corrected for Galactic extinction. The AT\,2022cmc spectra have been calibrated to the approximate light curve flux as described in Section \ref{sec:analysis}. Because the TDE-featureless spectra have not been host subtracted, we do not calibrate those to the transient light curve flux.}
    \label{fig:allspec}
\end{figure*}

\citet{Andreoni2022} presented 6 optical spectra of AT\,2022cmc taken over $\sim$16 days (observed frame), starting from $\sim$4 days after the first ZTF detection, which we reproduce here (Figure \ref{fig:allspec}). The collection of optical data includes spectra taken with NOT+ALFOSC, Gemini+Gemini Multi-Object Spectrograph (GMOS), Very Large Telescope (VLT) + Xshooter, Keck-II+DEep Imaging Multi-Object Spectrograph (DEIMOS), and Keck-I+Low Resolution Imaging Spectrometer (LRIS; \citealt{oke95}). We point the reader to \citet{Andreoni2022} for details on the data reduction. Several absorption lines in the VLT+Xshooter spectrum were identified as \ion{Al}{3}, \ion{Fe}{2}, \ion{Mn}{2}, \ion{Mg}{2}, \ion{Mg}{1}, and \ion{Ca}{2} and were used to obtain the source redshift of $z=1.193$.

We also present here for the first time an additional spectrum taken with Keck-I+LRIS (PI: Margutti) on 2022 Mar 31.62, $\sim 28$ days (observed frame) after the last spectrum published by \citet{Andreoni2022}. The 1\arcsec~slit was used with the 400/3400 grism on the blue side, along with the D560 dichroic and 400/8500 grating on the red side to cover the full optical range, with total exposure times of 1840/1800~s on the blue/red sides, respectively. The data reduction followed the standard procedures outlined by \citet{silverman12}.

The spectra cover both the red and blue phases of the optical light curve, and while the spectra remain featureless over all epochs, a clear evolution in the continuum is seen. In Figure \ref{fig:allspec}, we show the 6 spectra of AT\,2022cmc presented in \citet{Andreoni2022} and the new LRIS spectrum presented here. In Table \ref{tab:spec_22cmc_feat}, we summarize the spectroscopic observations of AT\,2022cmc.

\begin{deluxetable*}{ccccc}
\tablecaption{Summary of spectroscopic observations.}
\tablehead{ \colhead{Name} & \colhead{Redshift} & \colhead{Telescope+Inst.} & \colhead{Date} & \colhead{Phase (rest-frame days)}}
\startdata
\multirow{6}{*}{AT\,2022cmc} & \multirow{6}{*}{1.193} & NOT+ALFOSC & 2022 Feb 15 & 1.8\\
& & Gemini+GMOS & 2022 Feb 15 & 1.8\\
& & VLT+Xshooter & 2022 Feb 17 & 2.7\\
& & Keck-II+DEIMOS & 2022 Feb 17 & 2.7\\
& & Keck-I+LRIS & 2022 Feb 25 & 6.4\\
& & Keck-I+LRIS & 2022 Mar 03 & 9.1\\
& & Keck-I+LRIS & 2022 Mar 31 & 21.9\\
\hline
AT\,2018jbv & 0.340 & LDT+DeVeny & 2019 Mar 28 & 75.9 \\
AT\,2020qhs & 0.345 & LDT+DeVeny & 2020 Oct 11 & 57.3 \\
AT\,2020riz & 0.435 & LDT+DeVeny & 2020 Oct 15 & 39.8 \\
AT\,2020ysg & 0.277 & LDT+DeVeny & 2021 Jan 11 & 67.4
\enddata
\tablecomments{Summary of spectroscopic observations for AT\,2022cmc and the four TDE-featureless objects in \citet{Hammerstein23}. The first six spectra of AT\,2022cmc are from \citet{Andreoni2022} and we present an additional late spectrum here. The TDE-featureless objects are taken at much later phases due to the late classification. We include here the redshift, telescope and instrument used, date the spectrum was taken, and the rest-frame phase (i.e., time relative to flare peak that the observation was taken\tablenotemark{a}).}
\label{tab:spec_22cmc_feat}
\tablenotetext{a}{We use time since first detection for AT\,2022cmc.}
\end{deluxetable*}

\section{Analysis} \label{sec:analysis}
Throughout this paper, we compare the light curve of AT\,2022cmc to the light curves of non-jetted TDEs. We draw our comparison sample and light curve data from \citet{Hammerstein23} and use the light curve fit parameters presented there. We also fit the spectra of the four TDE-featureless objects in the \citet{Hammerstein23} sample, using the spectra presented in the Appendix of that paper, all of which were observed with the DeVeny spectrograph on the LDT. In Figure \ref{fig:allspec}, we show the optical spectra of the 4 TDE-featureless objects presented in \citet{Hammerstein23}. We describe the methods used to fit the spectra and light curves in the follow sections.

\subsection{Spectrum Fitting} \label{sec:specfit}
Given the faintness of the host galaxy \citep[$\gtrsim 24.5$ mag,][]{Andreoni2022} and the luminosity of the flare, we expect the optical spectra to be dominated by transient light. It is therefore of interest to characterize the properties of the optical spectra to gain information on the transient itself. While the optical spectra are primarily featureless, the continuum may provide information on the evolution of thermal and non-thermal emission.

Before we perform continuum fitting, we first scale the spectra by the light curve flux interpolated to the epoch of a particular spectrum. To do this, we simulate $g$-, $r$-, $i$-, and $z$-band observations from the spectra using \texttt{pysynphot} and fit the ratio between this simulated flux and the light curve flux with a first-order polynomial in order to get a wavelength-dependent correction factor. We note that in some cases (e.g., the DEIMOS spectrum) the spectral coverage only matched the $r$- and $i$-bands. In contrast to AT\,2022cmc, the hosts for the four TDE-featureless events are all detected and may influence the continuum shapes in the spectra. To account for this, we fit these spectra as a linear combination of the model and the host spectrum obtained with stellar population synthesis fits performed in \citet{Hammerstein23}. All spectra have been corrected for Galactic extinction prior to fitting. The spectra of the four featureless TDEs have been corrected for telluric absorption features, while only the Keck+LRIS spectra of AT\,2022cmc have been corrected for telluric features.

We fit the continuum of all AT\,2022cmc spectra and the four featureless TDEs using two different models. The first model is a blackbody which allows us to obtain the blackbody temperature and inferred blackbody radius, with the temperature bounds of [$10^4$, $10^5$] K, as is expected for TDEs \citep[e.g.,][]{vanVelzen21, Hammerstein23}. The second model is a power-law of the form:
\begin{equation}
    \label{eq:PL-Ch5}
    F_\nu = F_{\rm \nu. pl} \left ( \frac{\nu}{\nu_0} \right )^\alpha,
\end{equation}
where $F_\nu$ is the spectral flux density, $F_{\rm pl}$ is the reference spectral flux density at $\nu_0 = 10^{15}$ Hz, and $\alpha$ is the spectral index. We limit the value of $\alpha$ to [$-5$, 5] to account for the change in the continuum. Despite the lack of emission and absorption lines in the spectra, apart from some host galaxy stellar absorption lines, we mask regions commonly associated with emission in TDEs (see Section \ref{sec:intro}) and host stellar absorption lines during the continuum fits in addition to any remaining telluric absorption lines.

For comparison, we also fit the SED from the light curve, which has $ugriz$ coverage beyond what can be supplied by spectroscopy. We fit the same models (a simple power-law and blackbody) to bins of 1 day (in the rest-frame) which have observations in four unique bands. We describe the results and present the fits in Section \ref{sec:results}.

\subsection{Light Curve Fitting} \label{sec:lcfit}
For consistency with \citet{Andreoni2022}, we first fit the AT\,2022cmc light curve with a time-evolving power-law spectral component and a \textit{static} blackbody spectral component. In this model, the temperature and luminosity of the blackbody component are constant with time. We note that this model is unlikely to be able to reproduce the light curve at later times once the power-law component has faded. The model is described by:
\begin{equation}
    \label{eq:PL_statBB}
    L_\nu(t) = L_{\nu_0,\rm{peak}} \frac{B_\nu (T_0)}{B_{\nu_0}(T_0)} + L_{\rm \nu, pl} \times
     \begin{cases}
     10^{\beta_{\rm rise} (t-t_{\rm peak})} & t \leq t_{\rm{peak}}\\
     10^{\beta_{\rm decay} (t-t_{\rm peak})} & t > t_{\rm peak}
     \end{cases},
\end{equation}
where $L_{\rm \nu, pl}$ describes the power-law component luminosity and is a function of the power-law spectral flux density described by Equation \ref{eq:PL-Ch5} with power-law rise and decay timescales $\beta_{\rm rise}$ and $\beta_{\rm decay}$. Here $\nu_0$ refers to the reference frequency, which we have chosen to be $10^{15}$ Hz, and thus $L_{\nu_0,\rm{peak}}$ is the luminosity at the time of peak at this frequency. Because \citet{Andreoni2022} find little evolution of the power-law spectral index with time, and because we are primarily interested in the properties of the thermal component, we forgo performing a fit with a time-variable power-law spectral index. We limit this fit to the same data as presented in \citet{Andreoni2022} instead of using the entire light curve. We will refer to this fit as Model 1.

To compare the UV/optical light curve of AT\,2022cmc to the thermal TDEs, we adopt fitting methods similar to \citet{vanVelzen21} and \citet{Hammerstein23}. We now include a time-dependent blackbody spectral component which is fit with a Gaussian rise and an exponential decay. This model is described by:
\begin{equation}
    \label{eq:PLBB}
    \begin{aligned}
    L_\nu(t) = & L_{\rm \nu, pl} \times \begin{cases}
     10^{\beta_{\rm rise} (t-t_{\rm peak})} & t \leq t_{\rm{peak}}\\
     10^{\beta_{\rm decay} (t-t_{\rm peak})} & t > t_{\rm peak}
     \end{cases} \\ 
     &+ L_{\nu_0,\rm{peak}} \frac{B_\nu (T_0)}{B_{\nu_0}(T_0)}\times
     \begin{cases}
     e^{-(t-t_{\rm{peak}})^2/2\sigma^2} & t \leq t_{\rm{peak}}\\
     e^{-(t-t_{\rm peak})/\tau} & t > t_{\rm peak}
     \end{cases}.
     \end{aligned}
\end{equation}
This model fits for only one temperature, $T_0$, which is used to predict the luminosity in the other bands as a fraction of the luminosity at the reference frequency. We will refer to this model as Model 2.

Lastly, to allow for the most freedom in the evolution of the blackbody component, we perform a third fit similar to Model 2 but with a non-parametric temperature evolution similar to \citet{vanVelzen21} and \citet{Hammerstein23}. In this nonparametric temperature model, we fit the temperature at grid points spaced 10 days apart, throwing out grid points without data within 10 days of the point. As UV coverage is needed to strongly constrain the blackbody peak and is sparse for AT\,2022cmc, uncertainties in the temperature for grid points without UV coverage are larger. We will refer to this model as Model 3.

To estimate the parameters of the models above we use the \texttt{emcee} sampler \citep{Foreman-Mackey2013} using a Gaussian likelihood function that includes a ``white noise'' term, ln($f$), which accounts for any variance in the data not captured by the reported uncertainties and flat priors for all parameters. We use 100 walkers and 4000 steps, discarding the first 2000 steps to ensure convergence. The free parameters of the models are listed in Table \ref{tab:priors-Ch5}. We have chosen the priors for the power-law component to be consistent with \citet{Andreoni2022} but have modified several priors from \citet{Hammerstein23} for the blackbody component light curve timescales as we expect them to differ slightly. In summary, we perform the following fits to the AT\,2022cmc UV/optical light curve:
\begin{itemize}[itemindent=2em]
    \item[Model 1]: A fit with a time-dependent power-law component (with fixed $\alpha$) and a static blackbody component described by Equation \ref{eq:PL_statBB}.
    \item[Model 2]: A fit with a time-dependent power-law (with fixed $\alpha$) and a time-dependent blackbody (with constant temperature) component described by Equation \ref{eq:PLBB}.
    \item[Model 3]: A fit with a time-dependent power-law (with fixed $\alpha$) and a time-dependent blackbody (with non-parametric temperature evolution) component described by Equation \ref{eq:PLBB}.
\end{itemize}
We present the results of this light curve analysis in Section \ref{sec:results}.

We note the steep decline in the light curve at $\sim60$ days post-peak in the rest-frame. To quantify this, we fit a simple broken power-law to the $t\geq20$ day light curve in both the $g$- and $r$- bands using \texttt{emcee}. The broken power-law is of the form:
\begin{equation}
    L_\nu(t) = \begin{cases}
     L_{\rm \nu,break}\times(t/t_{\rm break})^{-\alpha_1} & t \leq t_{\rm{break}}\\
     L_{\rm \nu,break}\times(t/t_{\rm break})^{-\alpha_2} & t > t_{\rm{break}}
     \end{cases},
\end{equation}
where $L_{\rm \nu,break}$ is the luminosity in the specified band at the break time, $t_{\rm break}$, and $\alpha_1$ and $\alpha_2$ are the power-law indices before and after the break, respectively. We present the results of this fitting in Section \ref{sec:results}.

\begin{deluxetable*}{ccc}
\tablecaption{Light curve fit parameters and priors.}
\tablehead{\colhead{Parameter} & \colhead{Description} & \colhead{Prior}}
\startdata
$\log L_{\rm BB,\nu_0}$ & Peak blackbody luminosity & [$L_{\rm max}/2$, $2L_{\rm max}$] \\
$\log(T_0/{\rm K})$ & Mean temperature & [4, 5] \\
$t_{\rm peak, BB}$ & Time of peak & [$-2$, $20$] days \\
$\log(\sigma_{\rm BB}/\rm{day})$ & Gaussian rise time & [0, 1.5]\\
$\log(\tau_{\rm BB}/\rm{day})$ & Exponential decay time & [0, 3]\\
\hline
$\log(F_{\rm peak,pl}/\rm{ erg~s^{-1}~Hz^{-1}})$ & Peak power-law flux density & [10, 80] \\
$\alpha$ & Power-law spectral index & [$-$10, 0] \\
$t_{\rm peak, pl}$ & Time of peak & [$-2$, $2$] days \\
$\beta_{\rm rise}$ & Power-law rise time & [0, 100] days \\
$\beta_{\rm decay}$ & Power-law decay time & [$-100$, 0] days \\
\hline
$\ln{f}$ & White noise factor & [$-$5, $-$1.8] \\
\enddata
\tablecomments{The free parameters and corresponding priors for the light curve analysis described in Section \ref{sec:lcfit}. $L_{\rm max}$ is the observed maximum luminosity in any band. The table is split into blackbody parameters (top section ) and power-law parameters (middle section).}
\label{tab:priors-Ch5}
\end{deluxetable*}

\section{Results} \label{sec:results}
\subsection{Spectrum Fitting}
\begin{figure*}
    \centering
      \setkeys{Gin}{width=\columnwidth}
        \begin{subfigure}{}
         \includegraphics{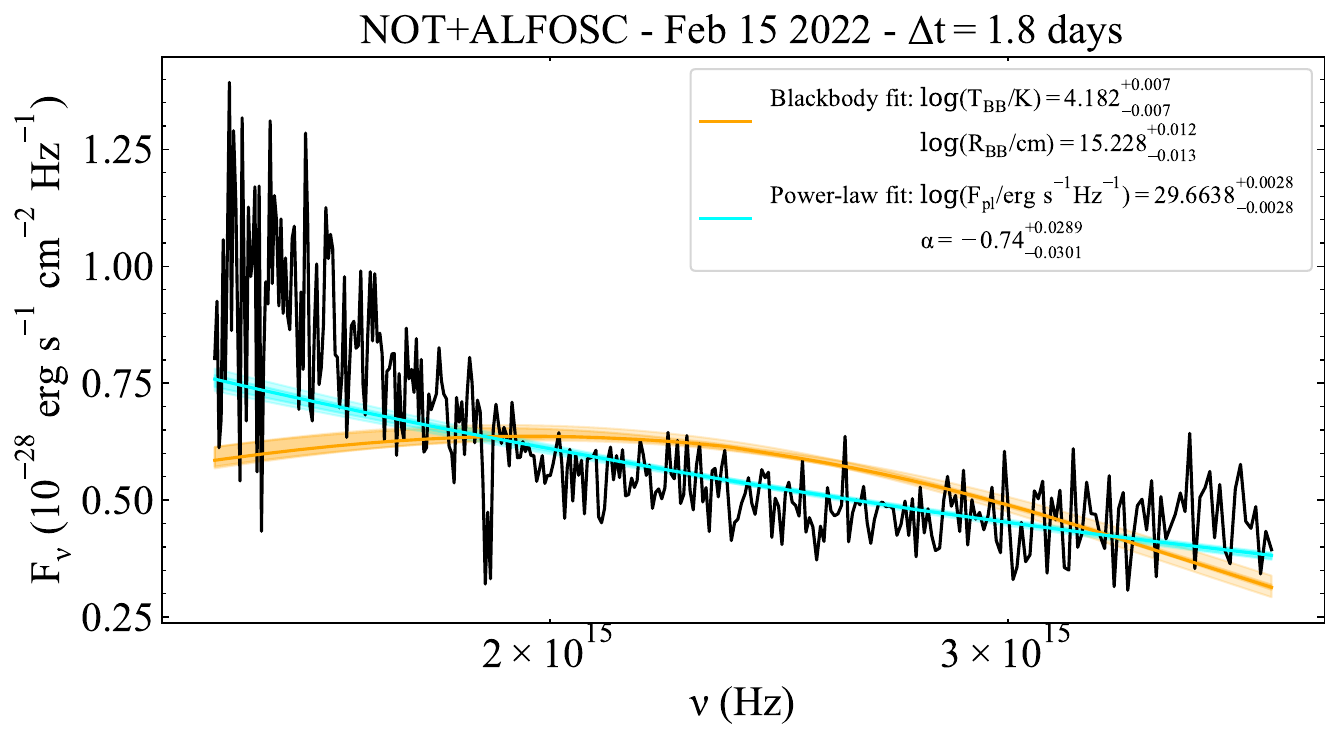}
        \end{subfigure}
        \begin{subfigure}{}
         \includegraphics{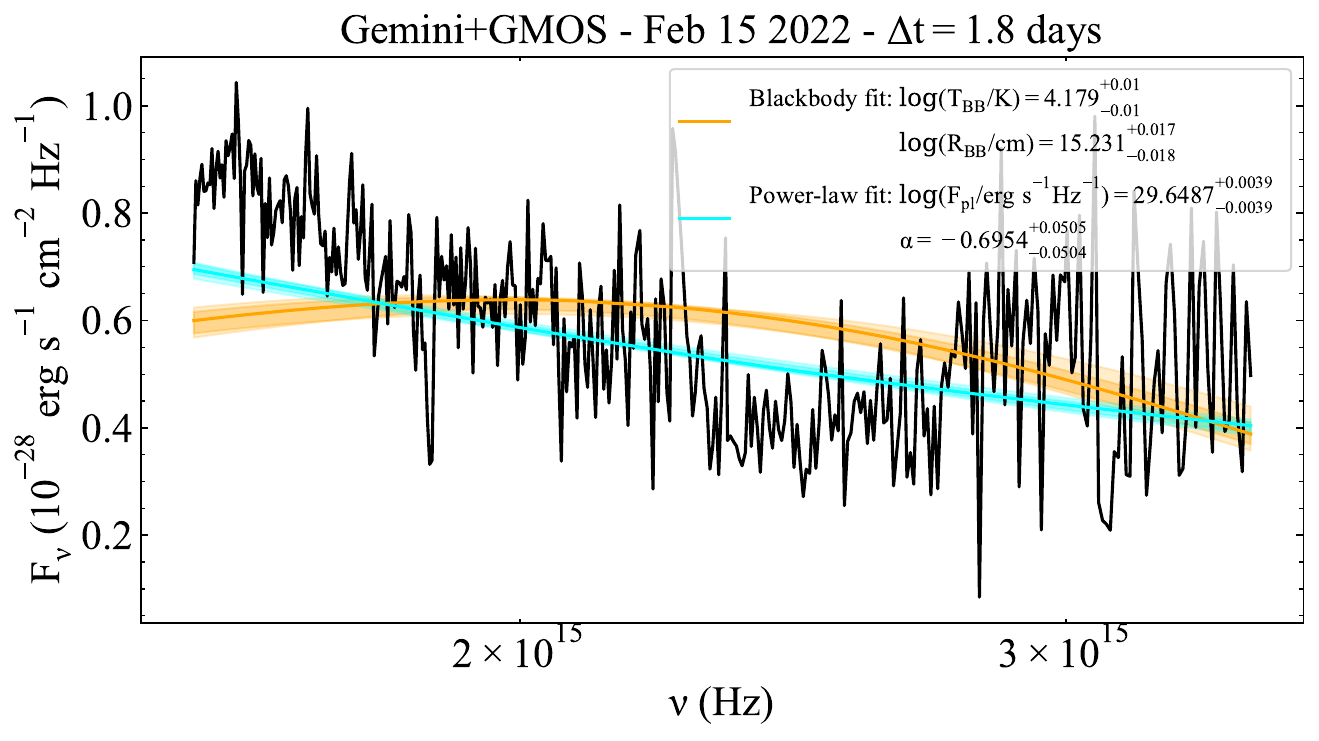}
        \end{subfigure}
        \begin{subfigure}{}
         \includegraphics{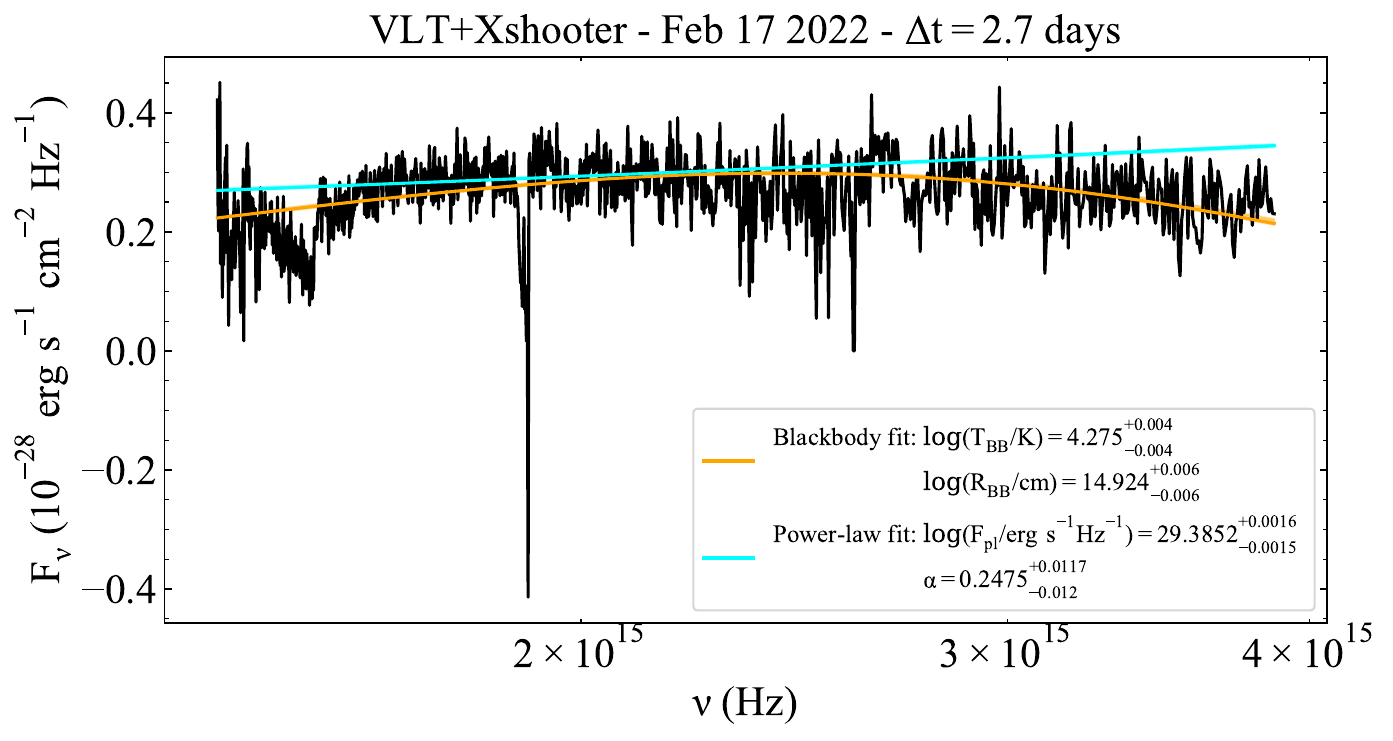}
        \end{subfigure}
        \begin{subfigure}{}
         \includegraphics{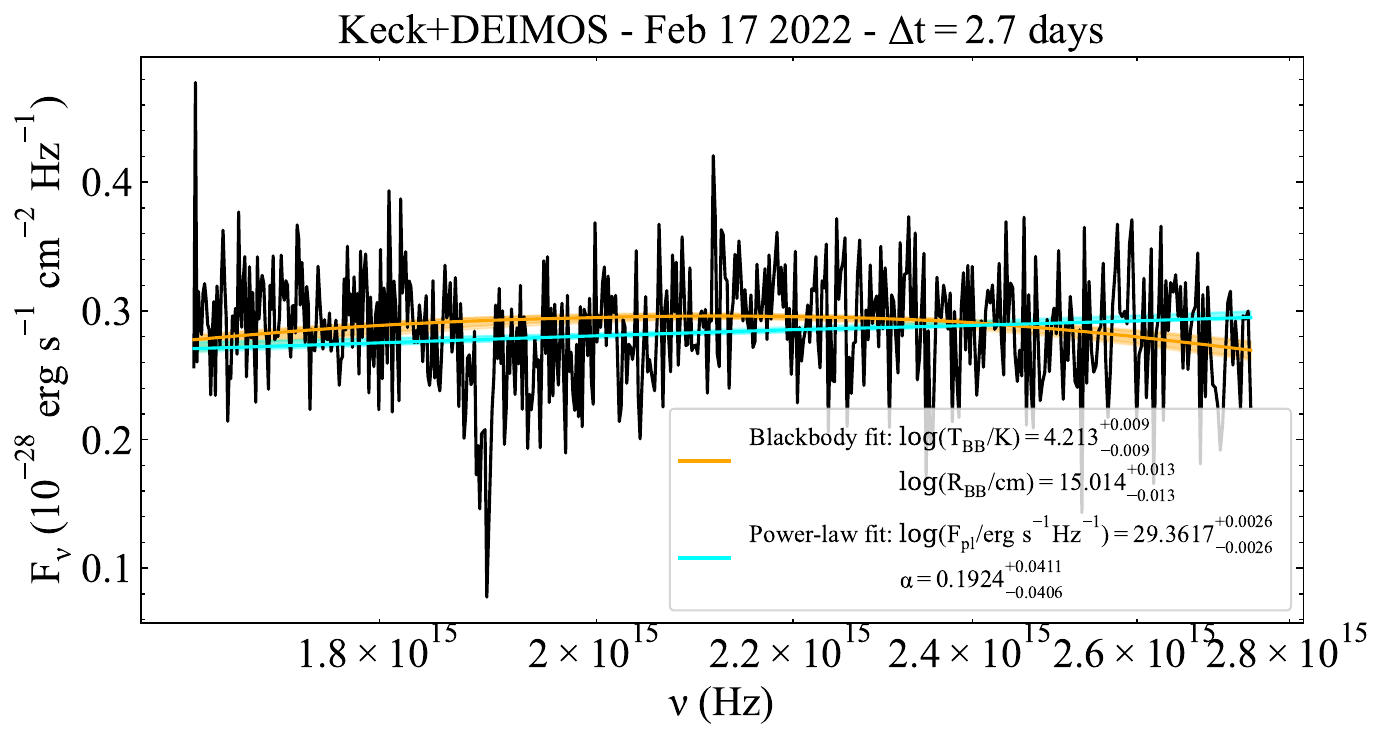}
        \end{subfigure}   
        \begin{subfigure}{}
         \includegraphics{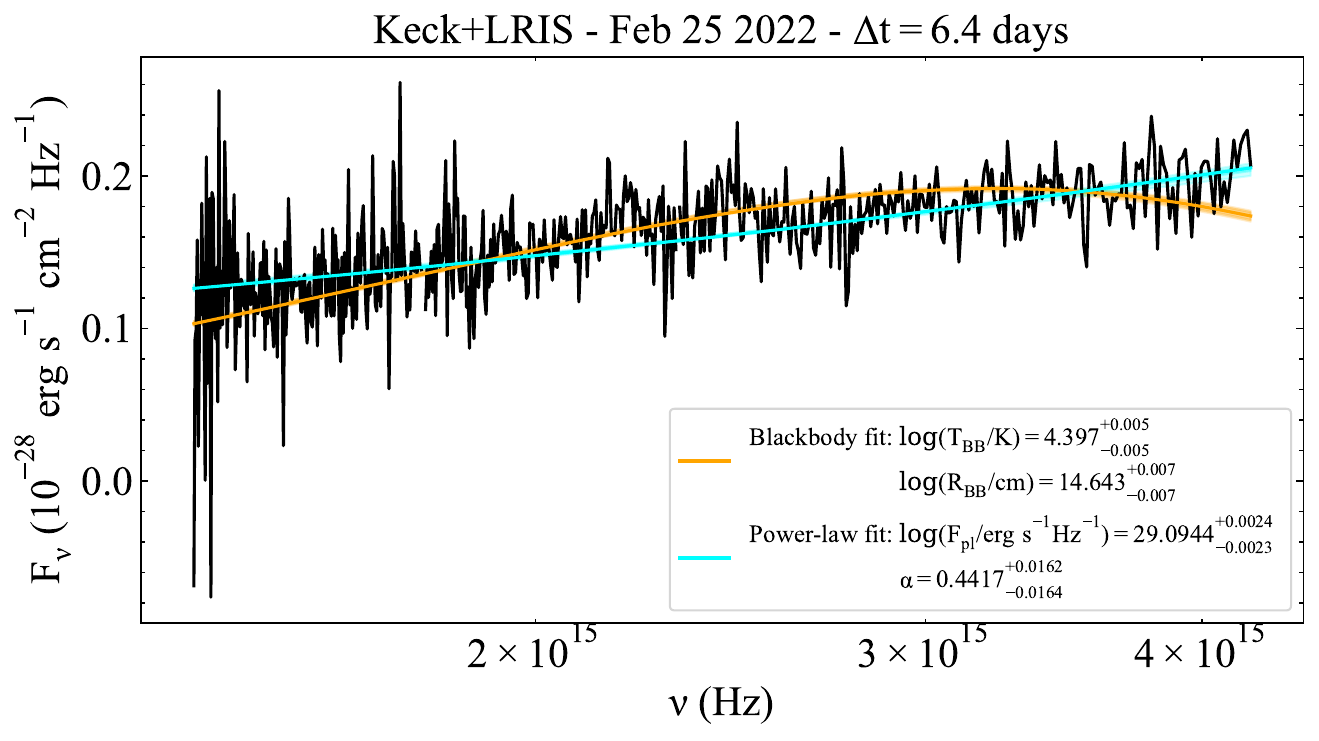}
        \end{subfigure}   
        \begin{subfigure}{}
         \includegraphics{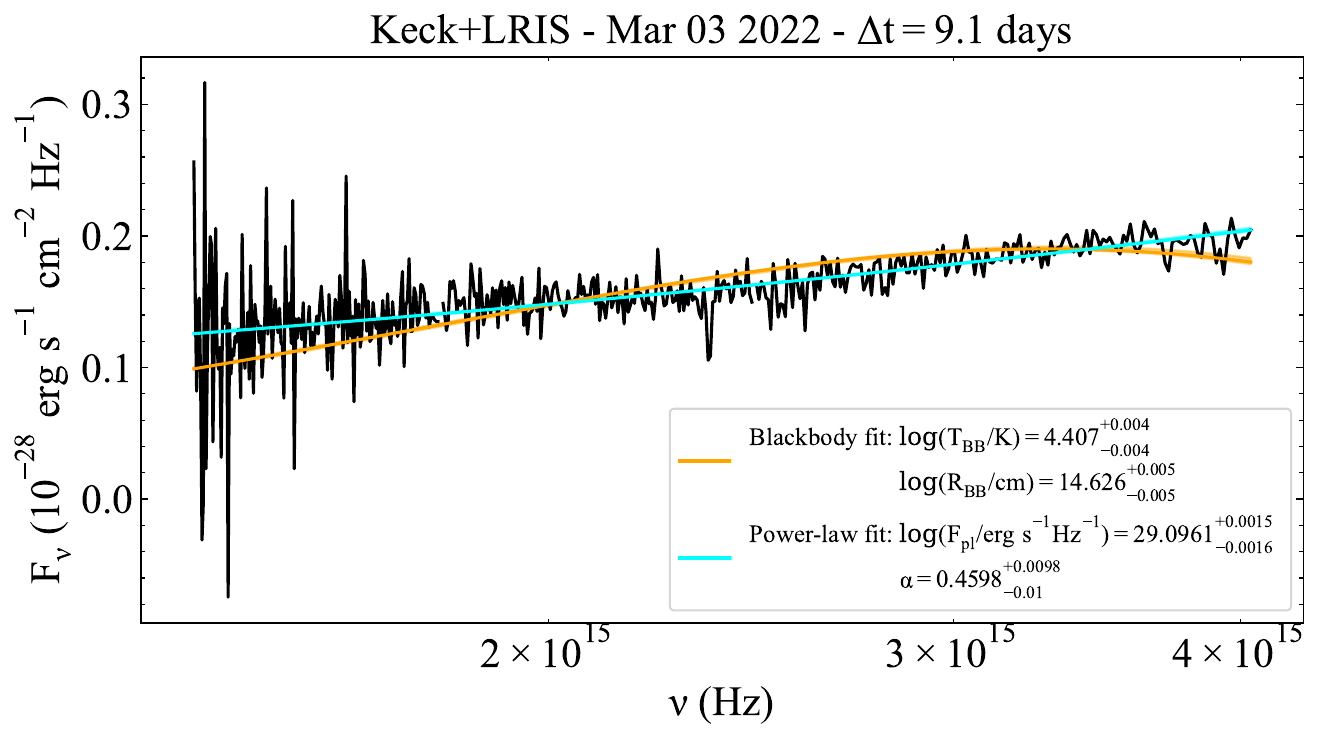}
        \end{subfigure}   
        \begin{subfigure}{}
         \includegraphics{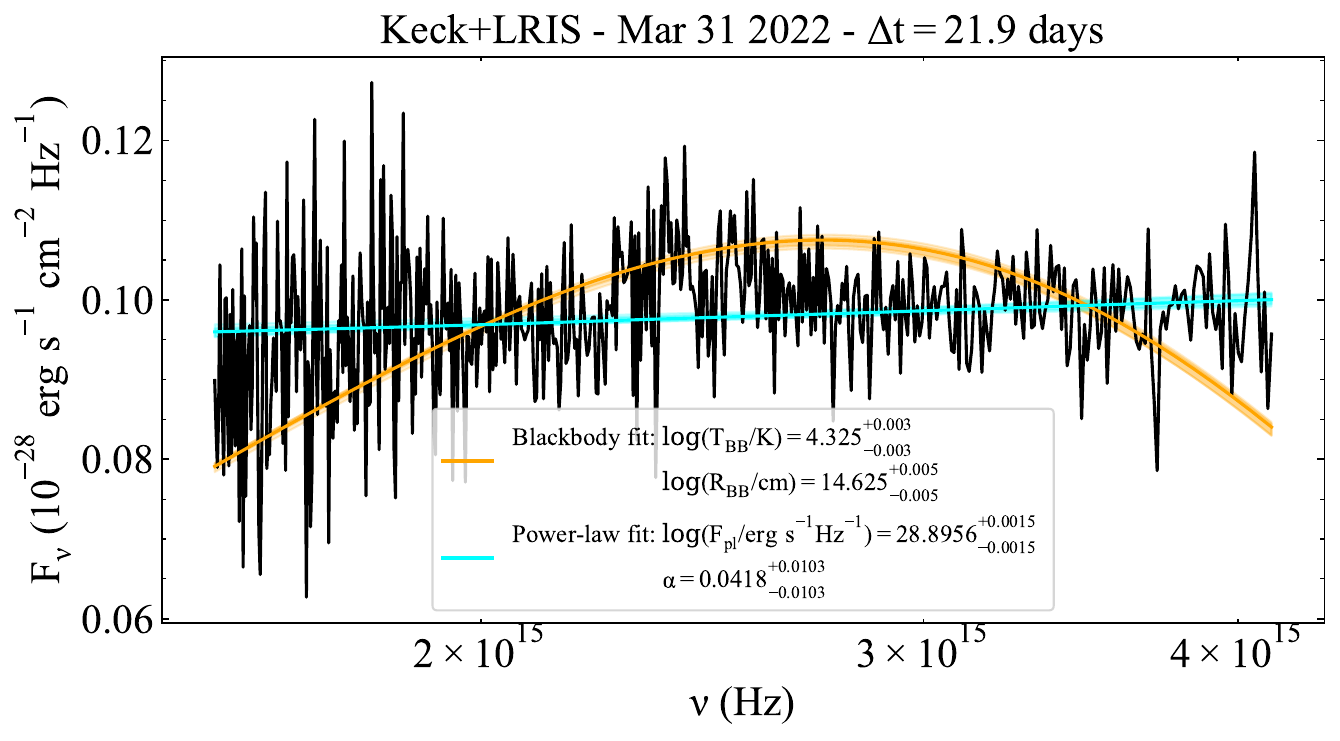}
        \end{subfigure}              

    \caption{The blackbody (orange line) and power-law (blue line) fits for AT\,2022cmc described in Section \ref{sec:specfit}. The early-time ($\Delta t \lesssim 2$ days) AT\,2022cmc spectra are well-described by a red power-law continuum, consistent with the red phase of the light curve, while the late-time ($\Delta t \gtrsim 6$ days) spectra are more consistent with a bluer continuum as expected in the blue phase of the light curve. The change in the continuum from 2022 Mar 03 to 2022 Mar 31 is also seen in the light curve, where the SED begins to redden once again and may either be intrinsic transient evolution or related to a greater fractional contribution from the host at later times. We note that for the later epochs, while the power-law may provide a better overall fit, the power-law spectral index is inconsistent with that of the red phase of the light curve (i.e., $\alpha > 0$ instead of $\alpha < 0$ in the red phase). The absorption features below $2\times10^{15}$ Hz are telluric features which are masked during fitting. Shading for each line corresponds to 1-$\sigma$ (darker shading) and 3-$\sigma$ (lighter shading) bands. All spectra are in the rest-frame. Fit parameters are listed in Table \ref{tab:specfit}.}
    \label{fig:specfits_cmc}
\end{figure*}

\begin{figure*}
    \centering
      \setkeys{Gin}{width=\columnwidth}
        \begin{subfigure}{}
         \includegraphics{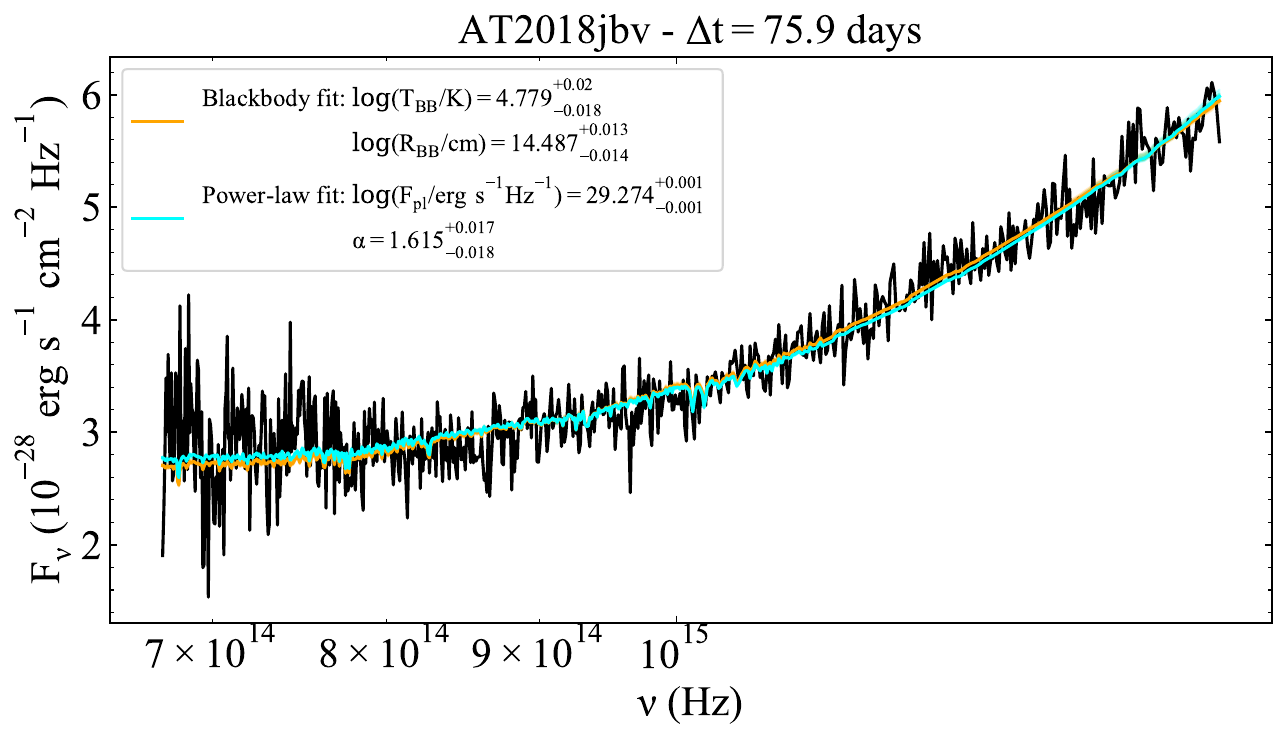}
        \end{subfigure}
        \begin{subfigure}{}
         \includegraphics{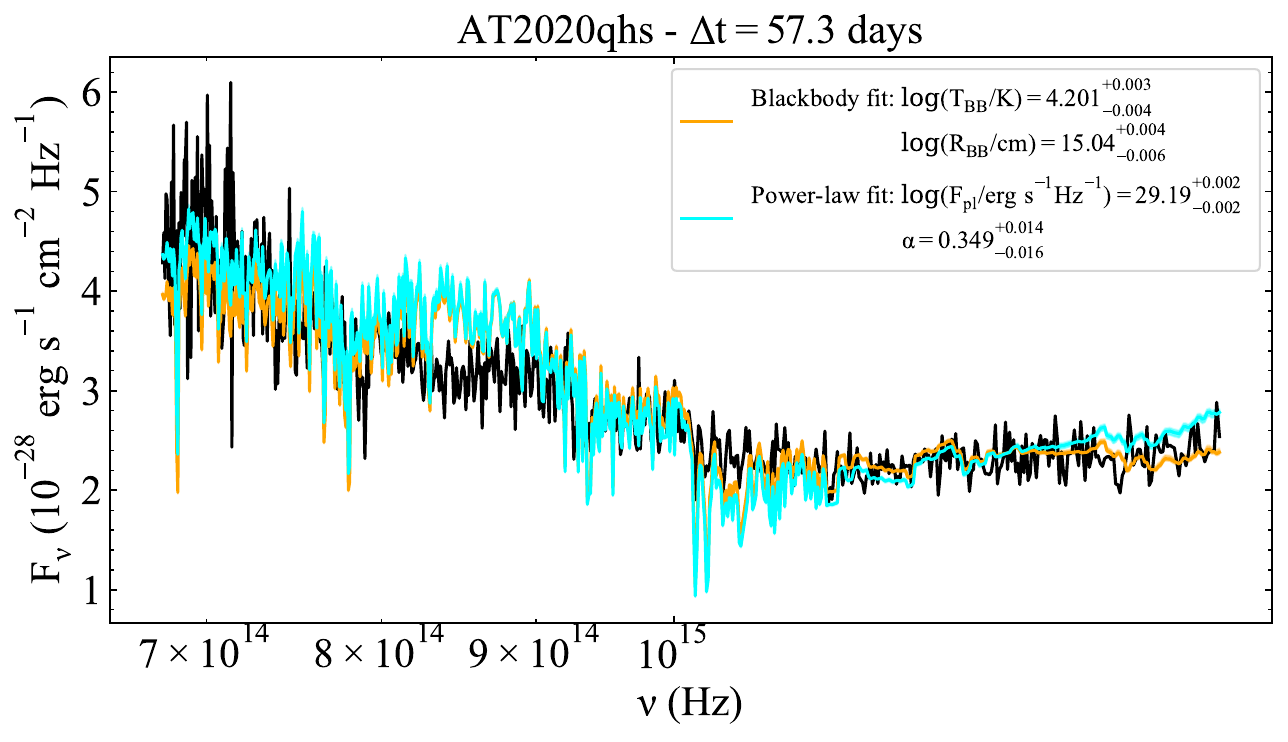}
        \end{subfigure}
        \begin{subfigure}{}
         \includegraphics{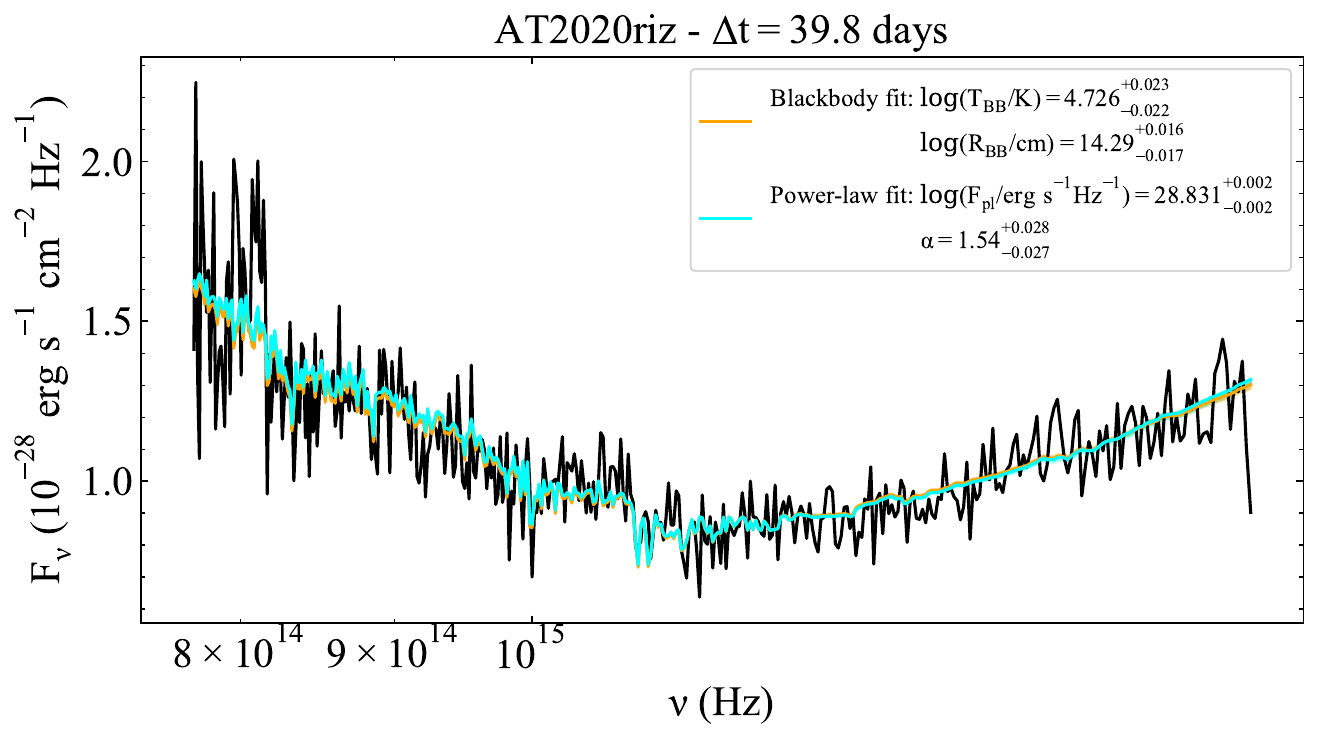}
        \end{subfigure}
        \begin{subfigure}{}
         \includegraphics{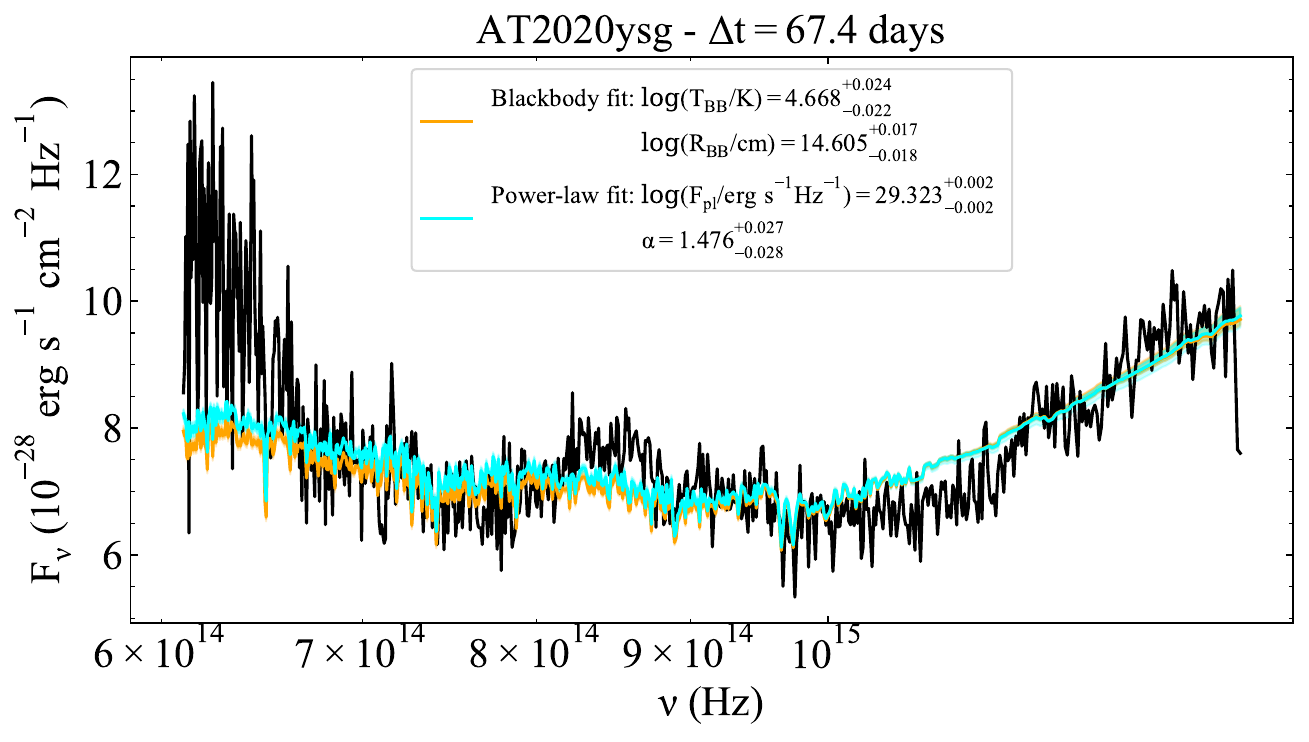}
        \end{subfigure}     

    \caption{The blackbody (orange line) and power-law (blue line) fits for the featureless TDEs from \citet{Hammerstein23}, including contribution from the host SED. The power-law fits are mostly inconsistent with the non-thermal emission that is seen for AT\,2022cmc, in that they correspond to positive $\alpha$ values (i.e., bluer) while early spectroscopy of AT\,2022cmc shows negative $\alpha$ values (i.e., redder). Nonetheless, the continuum shapes, particularly in AT\,2018jbv, are reminiscent of the later, blue phase AT\,2022cmc spectra. Shading for each line corresponds to 1-$\sigma$ (darker shading) and 3-$\sigma$ (lighter shading) bands. For AT\,2020ysg, we note the bump between $8-9
    \times 10^{14}$ Hz, which may correspond to the \ion{He}{2} $\lambda$3203\,\AA~line. All spectra are in the rest-frame. Fit parameters are listed in Table \ref{tab:specfit}.}
    \label{fig:specfits_feat}
\end{figure*}

In Figures \ref{fig:specfits_cmc} and \ref{fig:specfits_feat}, we show the results of the spectrum fitting described in \ref{sec:specfit}. For each spectrum, the power-law and blackbody fits to the continuum are shown, along with the fit parameters. We summarize these in Table \ref{tab:specfit}. The early-time spectra of AT\,2022cmc are generally not well-fit by the blackbody continuum model and in some cases produce temperatures lower and inferred radii larger than the sample of optically discovered thermal TDEs \citep[e.g.,][]{vanVelzen21, Hammerstein23, Yao2023}. The early-time emission is described much better by the power-law fits to the continuum which is unsurprising given that the red component dominates the light curve until $\sim 12$ days (observed frame) from the first detection \citep{Andreoni2022}. The blackbody continuum fits to the later spectra taken with LRIS, produce more reasonable results (i.e., those expected for TDEs) with regard to blackbody temperature and inferred radius. The spectra show remarkable similarity to the featureless TDEs (shown in the bottom 4 panels of Figure \ref{fig:specfits_cmc}).

\begin{figure}
    \centering
    \includegraphics[width=0.9\columnwidth]{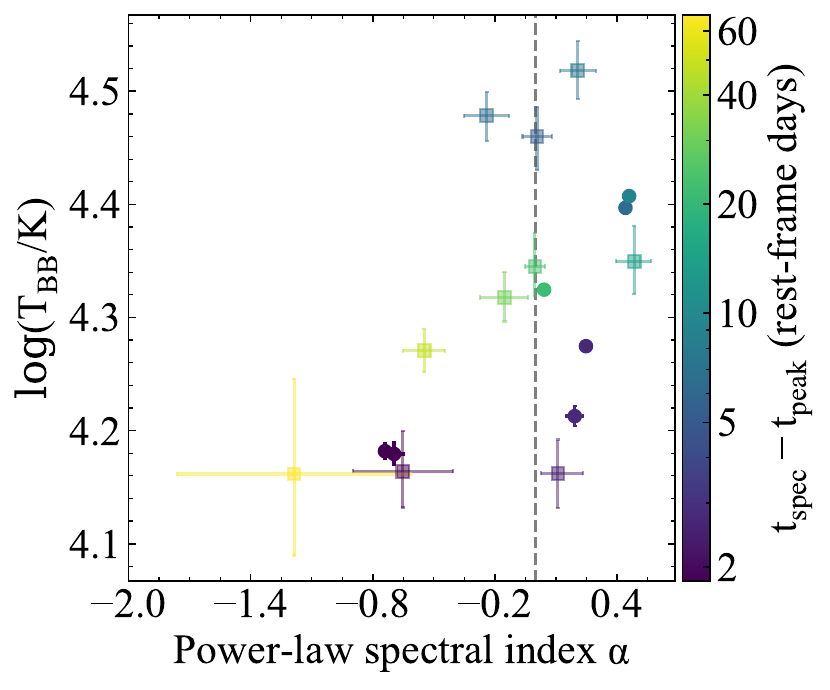}
    \caption{The evolution of the blackbody temperature $T_{\rm BB}$ and the power-law spectral index $\alpha$ from the spectrum continuum fits (circles). We also show the results from fitting the SED derived from the photometry (squares) There is a clear evolution in the power-law spectral index due to the fading of the red, non-thermal component in AT\,2022cmc after $\sim12$ days, where the blue, thermal component takes over. This is also evident in the fits to the SED. We note that the single model continuum fits (either power-law or blackbody) cannot fully capture the nature of the emission, but provide evidence for the clear evolution of the source emission in its red and blue phases. We denote $\alpha=0$ with the dashed gray line, negative and positive values of which roughly coincide with the red and blue phases of the light curves, respectively.}
    \label{fig:continuum_evo}
\end{figure}

While the single model continuum fits cannot completely capture the nature of the continuum emission, the change in the power-law spectral index $\alpha$ is evident from these fits. In Figure \ref{fig:continuum_evo}, we show the power-law spectral index $\alpha$ and the blackbody temperature for each spectrum as well as from the SED, which covers a larger range in time. The evolution from $\alpha = -0.74$ to $\alpha = 0.46$ is expected as the red, non-thermal component of the SED quickly fades in the first $\sim$10 days of the transient. Interestingly, the power-law index evolves back toward values less than zero after $\sim30$ days. Whether this is intrinsic evolution of the transient or is related to a larger fraction of the light being related to the host is difficult to determine without strong host constraints.

\begin{deluxetable*}{ccccccc}
\tablecaption{Results from spectrum fitting.}
\tablehead{Name & Telescope/Inst. & Phase (days) & $\log(F_{\rm pl}/{\rm erg~s^{-1}~Hz^{-1}})$ & $\alpha$ & $\log(T_{\rm BB}/{\rm K}$) & $\log(R_{\rm BB}/{\rm cm})$}
\startdata
\multirow{6}{*}{AT\,2022cmc} & NOT+ALFOSC & 1.8 & $29.664\pm0.003$ & $-0.74\pm0.03$ & $4.182\pm0.007$ & $15.228\pm0.012$\\
& Gemini+GMOS & 1.8 & $29.649\pm0.004$ & $-0.69\pm0.05$ & $4.179\pm0.01$ & $15.231\pm0.017$\\
& VLT+Xshooter & 2.7 & $29.385\pm0.002$ & $0.25\pm0.01$ & $4.275\pm0.004$ & $14.924\pm0.006$\\
& Keck+DEIMOS & 2.7 & $29.362\pm0.003$ & $0.19\pm0.04$ & $4.213\pm0.009$ & $15.014\pm0.013$ \\
& Keck+LRIS & 6.4 & $29.094\pm0.002$ & $0.44\pm0.02$ & $4.397\pm0.005$ & $14.643\pm0.007$\\
& Keck+LRIS & 9.1 & $29.096\pm0.002$ & $0.46\pm0.01$ & $4.407\pm0.004$ & $14.626\pm0.005$\\
& Keck+LRIS & 21.9 & $8.896\pm0.002$ & $0.04\pm0.01$ & $4.325\pm0.003$ & $14.625\pm0.005$\\
\hline
AT\,2018jbv & LDT+DeVeny & 75.9 & $29.274\pm0.001$ & $1.62\pm0.02$ & $4.779\pm0.020$ & $14.487\pm0.013$\\
AT\,2020qhs & LDT+DeVeny & 57.3 & $29.190\pm0.002$ & $0.35\pm0.01$ & $4.201\pm0.003$ & $15.040\pm0.004$\\
AT\,2020riz & LDT+DeVeny & 39.8 & $28.831\pm0.002$ & $1.54\pm0.03$ & $4.726\pm0.023$ & $14.290\pm0.016$\\
AT\,2020ysg & LDT+DeVeny & 67.4 & $29.323\pm0.002$ & $1.48\pm0.03$ & $4.668\pm0.024$ & $14.605\pm0.017$\\
\enddata
\tablecomments{Results from power-law and blackbody continuum fits to optical spectra. The power-law spectral index $\alpha$ for AT\,2022cmc begins negative and corresponds to the red component of the light curve in Figures \ref{fig:g-r} and \ref{fig:continuum_evo}. Notably, $\alpha$ begins to transition back toward zero in the latest epoch. This trend is seen in the late-time light curve epochs until $\sim60$ days. The nature of this evolution is unusual for a TDE and may be related to a greater fractional host component at later times.}
\label{tab:specfit}
\end{deluxetable*}

\subsection{Light Curve Fitting} \label{sec:lcresults}
\begin{figure}
    \centering
      \setkeys{Gin}{width=\columnwidth}
        \begin{subfigure}{}  % <----
         \includegraphics{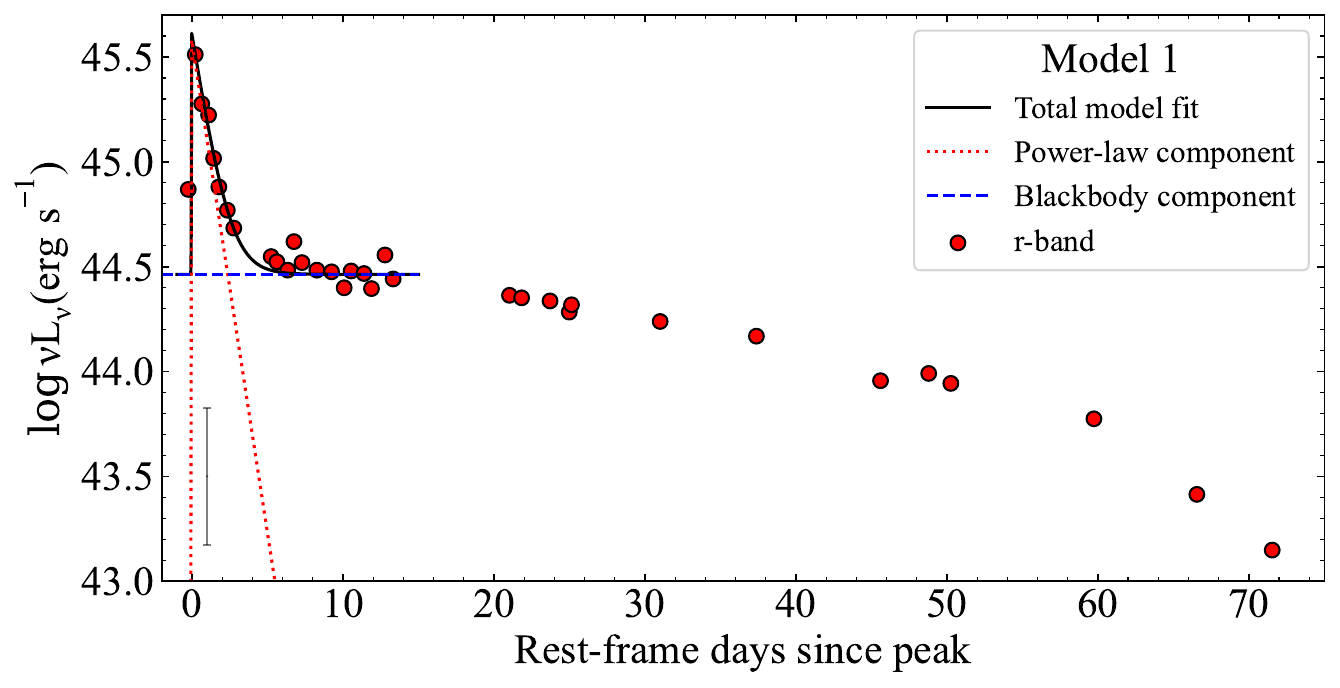}
        \end{subfigure}
        \vspace{-1.1em}
        \begin{subfigure}{}  % <----
         \includegraphics{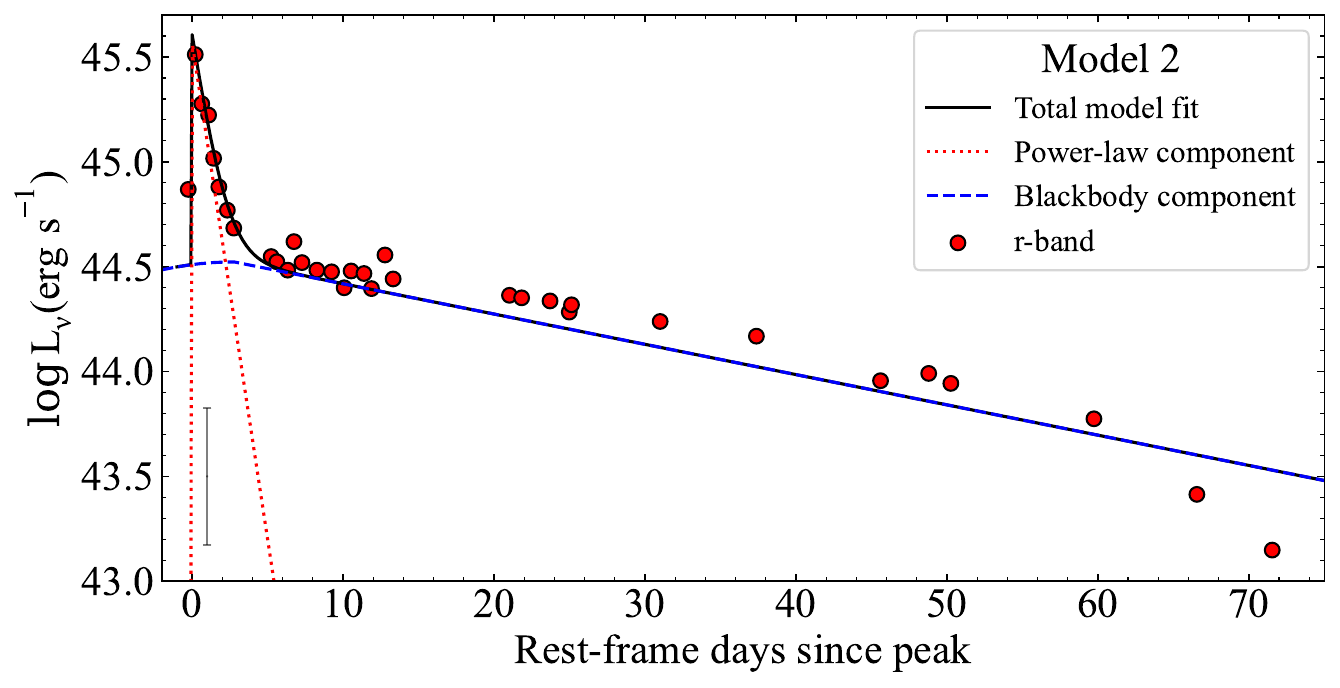}
        \end{subfigure}
        \vspace{-1.1em}
        \begin{subfigure}{}  % <----
         \includegraphics{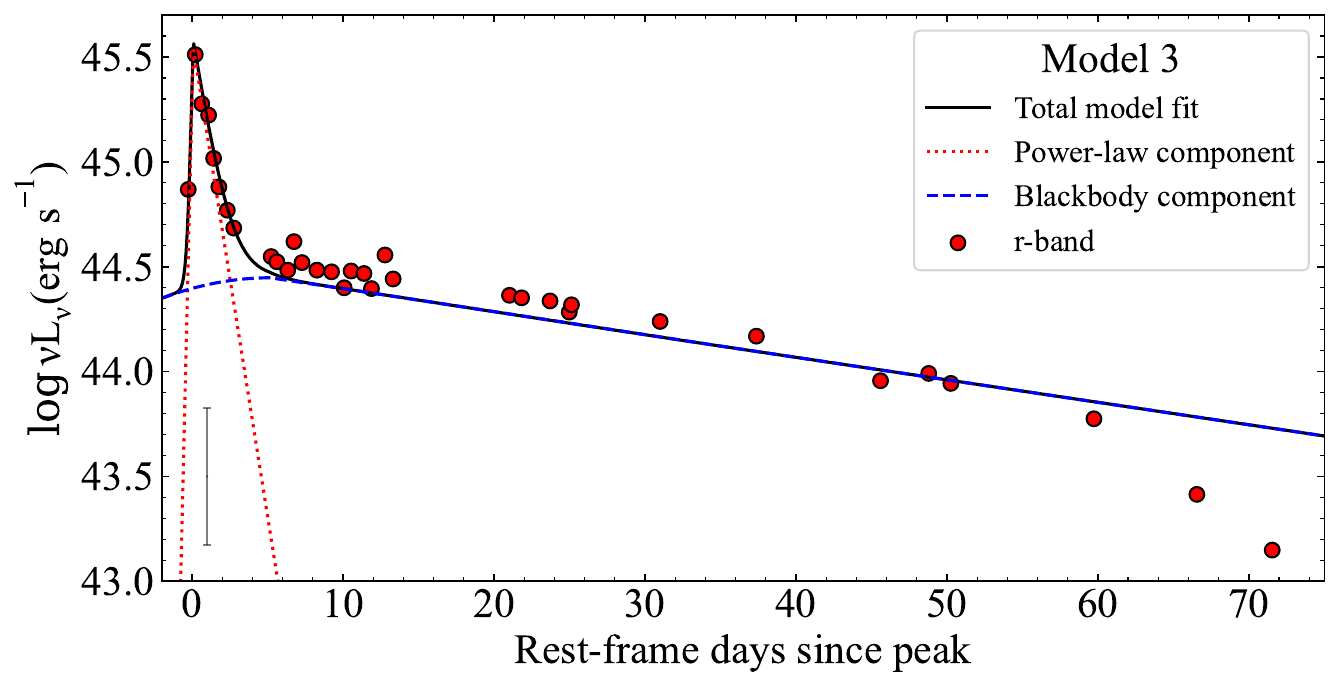}
        \end{subfigure}
    
    \caption{Model fits to the AT\,2022cmc light curve, showing only the $r$-band for clarity. For each fit we show the two components, the power-law and the blackbody as red dotted and blue dashed lines, respectively. The rise times for the two components, $\sigma_{\rm BB}$ and $\beta_{\rm rise}$, are derived from the red and blue curves. We note that the rise times, particularly for the power-law component, are not well-constrained. Additionally, we show the median uncertainties on the light curve data points in the bottom left corner. The Model 1 fit is limited to the first $\sim$15 days to match \citet{Andreoni2022}. We note the unusual accelerated decay after $\sim$60 days, which is uncommon in thermal TDEs.}
    \label{fig:lcfits_Ch5}
\end{figure}

In Table \ref{tab:lcfit_Ch5}, we show the results from fitting the three models described in Section \ref{sec:lcfit} to the UV/optical light curve of AT\,2022cmc. For the time-dependent power-law and static blackbody fit (Model 1), we find results consistent with the light curve fit performed by \citet{Andreoni2022}. While the blackbody luminosity we find is lower ($L_{\rm BB} = 10^{44.41}$ erg s$^{-1}$ compared to $L_{\rm BB} = 10^{45.53}$ erg s$^{-1}$) and the blackbody temperature is higher ($T_{\rm BB} = 10^{4.53}$ K compared to $T_{\rm BB} = 10^{4.48}$ K), our results are nonetheless consistent with typical values for TDEs. The time-dependent power-law component is consistent within uncertainties with the findings of \citet{Andreoni2022}, with the reference spectral flux density $F_{\rm peak, pl} = 10^{30.59}$ erg s$^{-1}$ Hz$^{-1}$ compared to their $F_{\rm peak, pl} = 10^{30.51}$ erg s$^{-1}$ Hz$^{-1}$, and the spectral index $\alpha = -1.30$ compared to their $\alpha = -1.32$\footnote{Note that \citet{Andreoni2022} label the power-law spectral index as $\beta$.}. When we allow the blackbody luminosity to vary with time, our results for both the power-law and blackbody components are consistent within mutual uncertainties. The Model 3 fits, where the temperature is allowed to evolve over the course of the light curve, produce only marginally different results, all but the power-law component rise time consistent within uncertainties. This is likely due to the lack of data on the rise of the light curve and the rapid nature of the rise. We show the light curve fits to only the $r$-band light curve (for clarity) in Figure \ref{fig:lcfits_Ch5}.

\begin{figure}
\centering
  \setkeys{Gin}{width=\columnwidth}
    \begin{subfigure}{}  % <----
     \includegraphics{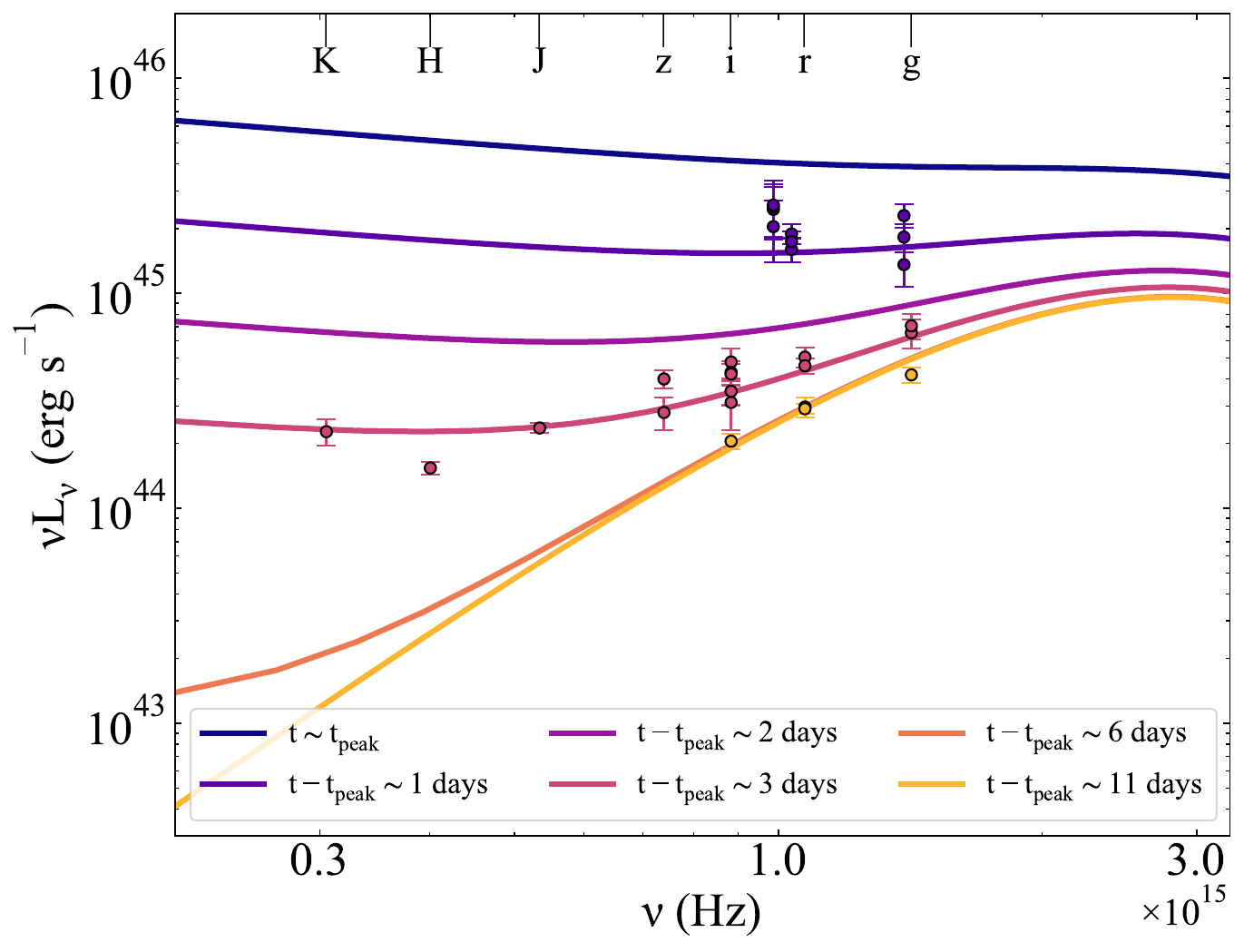}
    \end{subfigure}
    \vspace{-1.1em}
    \begin{subfigure}{}  % <----
     \includegraphics{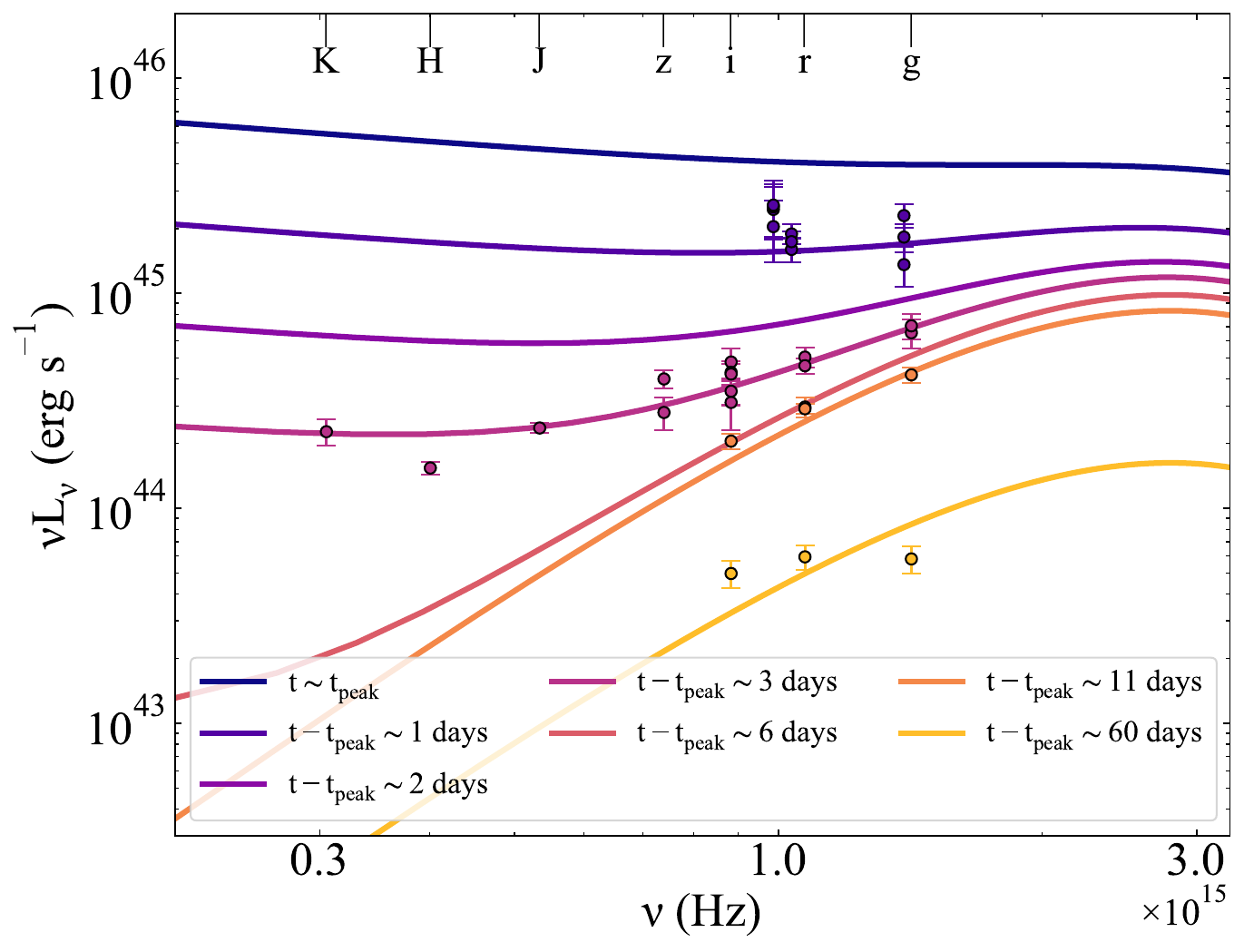}
    \end{subfigure}
    \vspace{-1.1em}
    \begin{subfigure}{}  % <----
     \includegraphics{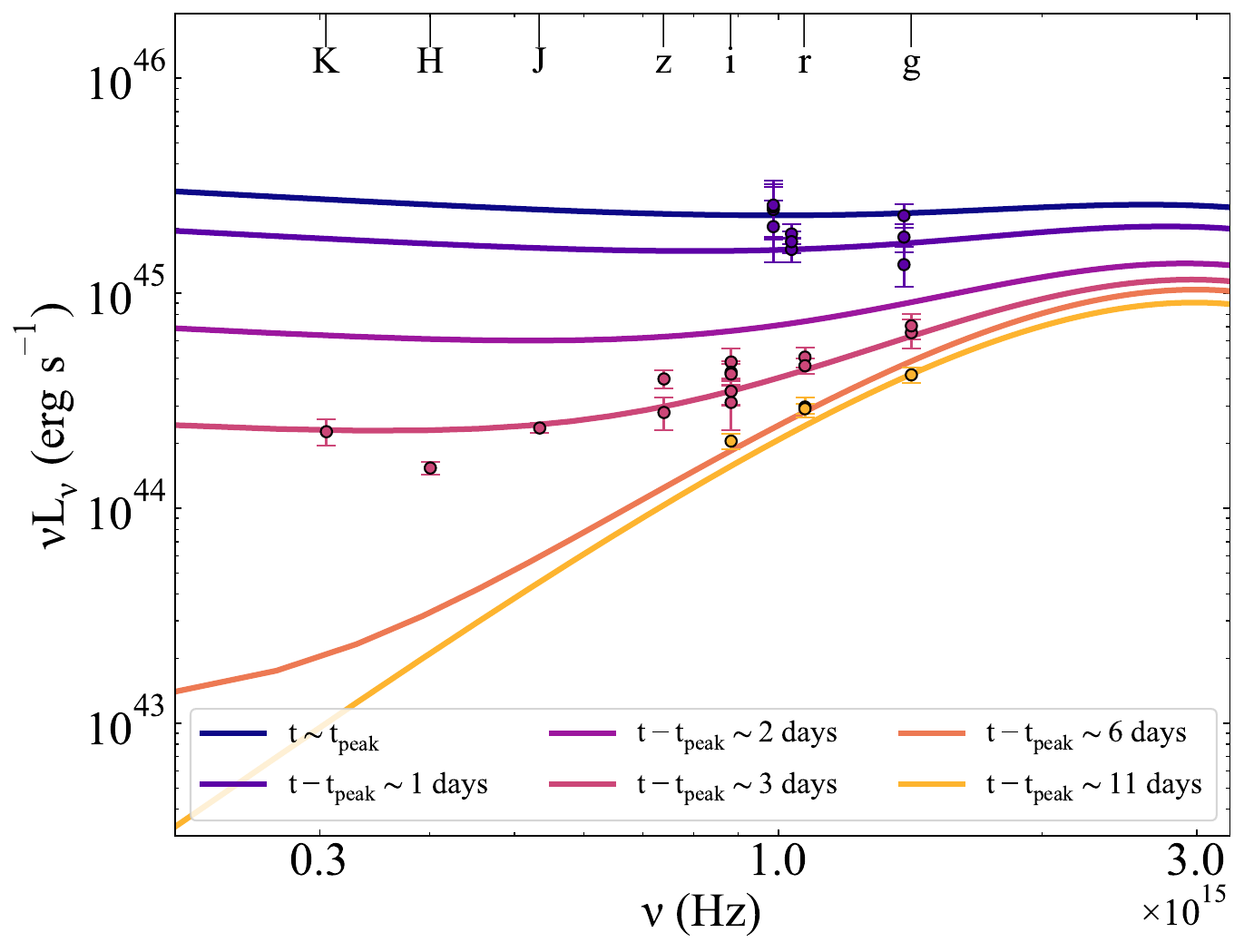}
    \end{subfigure}
    
\caption{Evolution of the SED starting from peak ($t = t_{\rm peak}$) until the $\sim$last epoch of observations (rest-frame) in order of Model 1 (top), Model 2 (middle), and Model 3 (bottom). For each panel, we show light curve data points for $t - t_{\rm peak} \approx 1, 3, 11, 60$ days. All SEDs are in the rest-frame.}
\label{fig:SED}
\end{figure}

In Figure \ref{fig:SED}, we show the SED derived from all fits, starting from the peak of the light curve until $\sim$60 days post-peak (rest-frame). There is a clear evolution in the shape of the SED in all cases, with the early-time SED dominated by the power-law component until $\sim6$ days post-peak. This is consistent with \citet{Andreoni2022}. The blackbody component is found to peak $\sim5$ days after the power-law component. Similar behavior is observed in the continuum in the optical spectra.

\begin{figure}
    \centering
    \includegraphics[width=0.9\columnwidth]{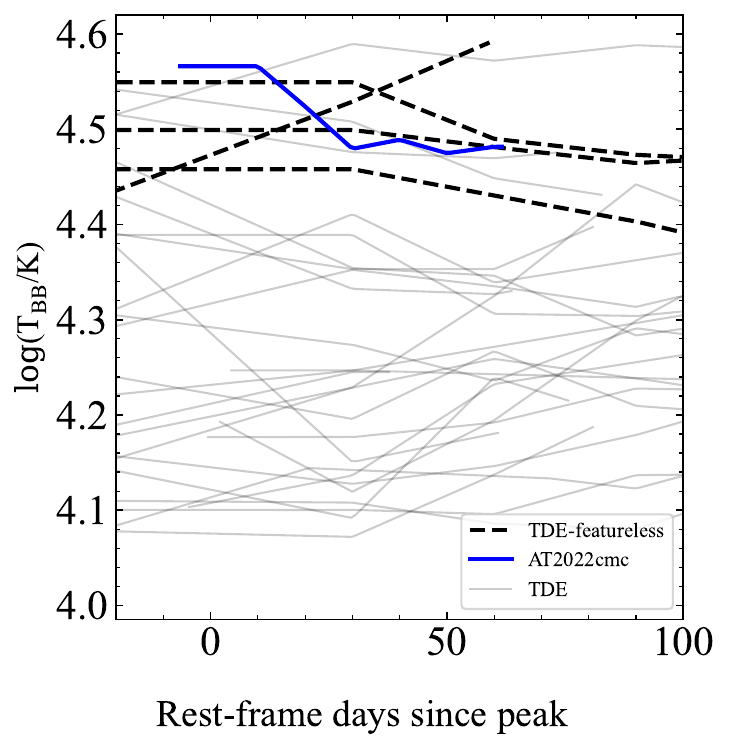}
    \caption{Evolution of the blackbody temperature from Model 3 with AT\,2022cmc represented by the solid blue line, the featureless TDEs represented by the dashed black lines, and other thermal TDEs as the gray solid lines. The thermal TDEs and the featureless TDEs show a variety of behaviors, although the majority show a near constant or increasing temperature \citep{Hammerstein23}. AT\,2022cmc appears to decrease in temperature, but the late-time temperature is not well constrained without UV observations.}
    \label{fig:tevol}
\end{figure}

In Figure \ref{fig:tevol}, we show the evolution of the blackbody temperature from Model 3. \citet{Hammerstein23} found that the blackbody temperature of the thermal TDEs is largely constant or even increasing, while the temperature of AT\,2022cmc appears to decrease. We note that there are very limited UV observations of AT\,2022cmc, which are needed to strongly constrain the peak of the blackbody SED. In the next section, we compare the blackbody component of the fit to other TDEs discovered in the optical.

In Table \ref{tab:brokenPL}, we present the results of the broken power-law fit to the $t\geq 20$ day light curve. There is generally good agreement between the $r$- and $g$- bands for both the break time and power-law indices. Both bands imply a break in the light curve at $\sim52$ days. We note the large uncertainties on the $\alpha_2$ power-law index, which is only constrained by a few data points.

\begin{deluxetable*}{cccc}
\tablecaption{Results from fits to AT\,2022cmc optical light curve.}
\tablehead{\colhead{Parameter} & \colhead{Model 1} & \colhead{Model 2} & \colhead{Model 3}}
\startdata
$\log (L_{\rm BB, \nu_0} /\rm{erg~s}^{-1})$ & $44.41_{-0.12}^{+0.09}$ & $44.46_{-0.14}^{+0.14}$ & $44.41_{-0.13}^{+0.12}$ \\
$t_{\rm peak, BB}$ (day)\tablenotemark{*} & -- & $2.78_{-3.36}^{+8.37}$ & $5.84_{-5.46}^{+7.80}$\\
$\log(\sigma_{\rm BB}/\rm{day})$ & -- & $1.06_{-0.58}^{+0.31}$ & $1.06_{-0.48}^{+0.31}$\\
$\log(\tau_{\rm BB}/\rm{day})$ & -- & $1.48_{-0.29}^{+0.48}$ & $1.53_{-0.40}^{+0.59}$\\
$\log(T_{\rm0}/\rm{K})$ & $4.53_{-0.06}^{+0.07}$ & $4.53_{-0.07}^{+0.08}$ & $4.57_{-0.27}^{+0.27}$\\
$\log(F_{\rm peak, pl}/\rm{erg~s^{-1}~Hz^{-1}})$ & $30.59_{-0.07}^{+0.07}$ & $30.59_{-0.07}^{+0.07}$ & $30.56_{-0.05}^{+0.06}$\\
$t_{\rm peak, PL}$ (day)\tablenotemark{*} & $-0.02_{-0.14}^{+0.15}$ & $-0.02_{-0.15}^{+0.14}$ & $0.03_{-0.14}^{+0.15}$\\
$\beta_{\rm rise}$ (day) & $50.43_{-35.01}^{+33.61}$ & $46.40_{-33.32}^{+34.34}$ & $3.62_{-1.72}^{+2.85}$\\
$\beta_{\rm decay}$ (day) & $-0.47_{-0.06}^{+0.05}$ & $-0.47_{-0.08}^{+0.06}$ & $-0.46_{-0.07}^{+0.06}$\\
$\alpha$ & $-1.30_{-0.41}^{+0.39}$ & $-1.31_{-0.51}^{+0.42}$ & $-1.11_{-0.44}^{+0.41}$\\
\enddata
\tablecomments{Results from fits to AT\,2022cmc optical light curve. The fit numbers correspond to those in Section \ref{sec:lcfit}.}
\tablenotetext{*}{Rest-frame, with respect to maximum light.}
\label{tab:lcfit_Ch5}
\end{deluxetable*}

\begin{deluxetable}{ccc}
\tablecaption{Results from broken power-law fits.}
\tablehead{\colhead{Parameter} & \colhead{$r$-band} & \colhead{$g$-band}}
\startdata
$\log L_{\rm \nu, break}$ & $43.92_{-0.31}^{+0.57}$ & $44.01_{-0.31}^{+0.52}$\\
$t_{\rm break}$ & $51.78_{-8.77}^{+11.02}$ & $51.87_{-8.51}^{+11.98}$ \\
$\alpha_1$ & $0.92_{-0.66}^{+1.21}$ & $1.00_{-0.71}^{+1.15}$ \\
$\alpha_2$ & $5.38_{-3.56}^{+3.20}$ & $5.13_{-3.58}^{+3.25}$
\enddata
\tablecomments{Results from broken power-law fits to AT\,2022cmc $r$- and $g$-band light curves. $t_{\rm break}$ is measured in the rest-frame with respect to the time of first detection. We note the large uncertainties on the $\alpha_2$ values, which are only constrained by a few data points.}
\label{tab:brokenPL}
\end{deluxetable}

\section{Discussion} \label{sec:discussion}
Several correlations have been found between properties of optical TDE light curves as well as between these properties and the SMBH mass (or host galaxy mass as a proxy). \citet{vanVelzen21} found that the rise time of the flare ($\sigma_{\rm BB}$) is correlated with the peak bolometric luminosity. This correlation was not found by \citet{Hammerstein23}, but their smaller sample size may affect this conclusion. \citet{Hammerstein23} did, however, find a weak correlation between the decay timescale and the peak luminosity, which has also been found by \citet{Hinkle2020}. \citet{vanVelzen21} also reported a positive correlation between the decay timescale ($\tau_{\rm BB}$) for their sample of TDEs and the host galaxy stellar mass. This correlation was also found in the \citet{Hammerstein23} sample, in addition to a correlation with the rise time, and a similar correlation was confirmed by \citet{Yao2023}. If the thermal, blue component in AT\,2022cmc arises in the same way as for thermal TDEs, we might expect that it falls in line with these same correlations. In this section, we compare the properties of the UV/optical emission in AT\,2022cmc to those of optically selected thermal TDEs, focusing on properties for which significant correlations have been reported. We use the sample of 30 TDEs from ZTF for comparison \citep{Hammerstein23}.

\begin{figure*}
    \centering
    \includegraphics[width=0.6\textwidth]{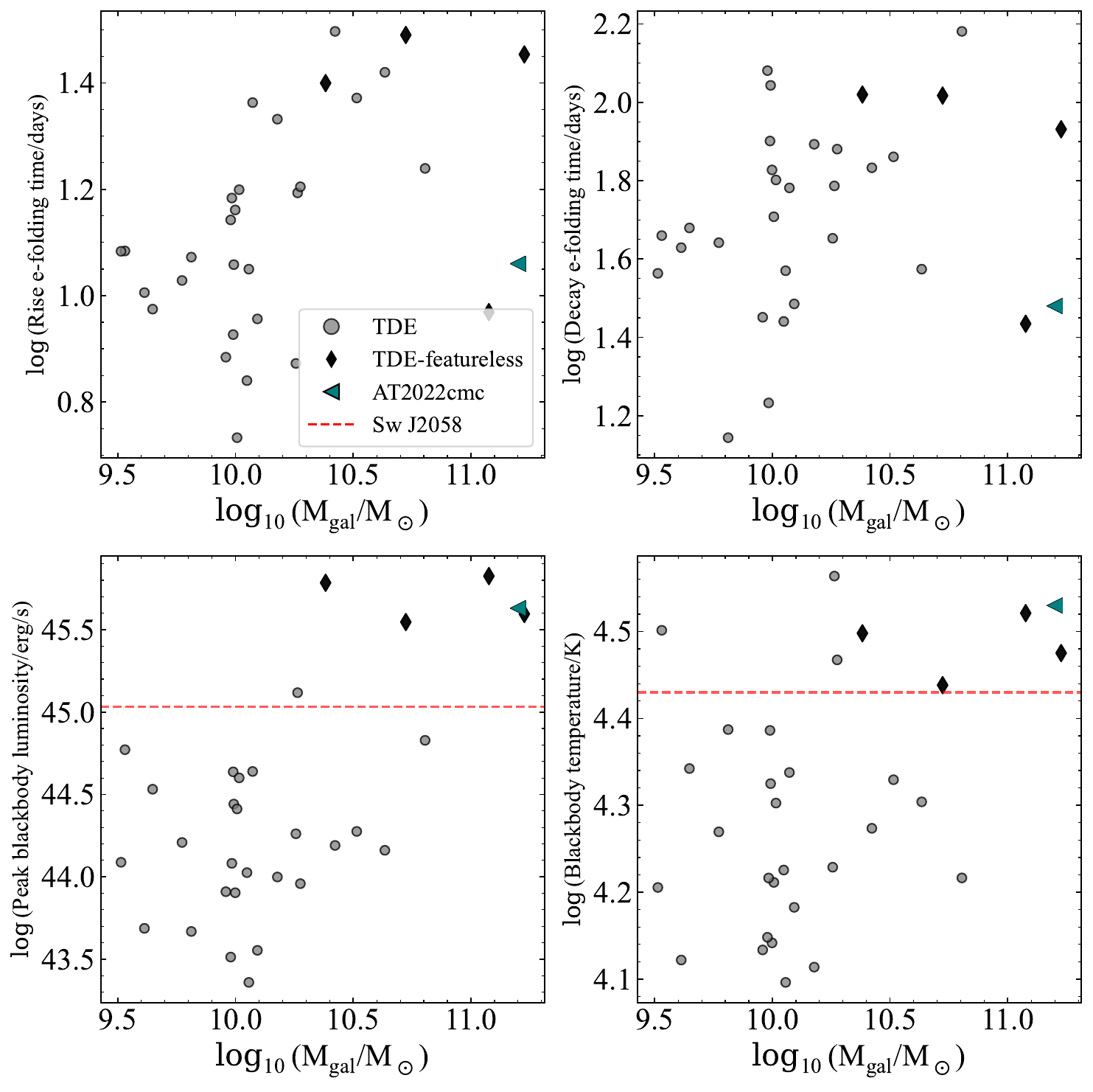}
    \caption{Selected light curve parameters from Model 1 for the 30 TDEs from ZTF \citep[gray circles, black diamonds indicate TDE-featurless;][]{Hammerstein23} and AT\,2022cmc (blue triangle). The rise and decay times for the blackbody component are much shorter for AT\,2022cmc than for the other TDEs. The peak bolometric blackbody luminosity and mean blackbody temperature are more consistent with expected values from TDEs and show remarkable similarity to the TDE-featureless class.}
    \label{fig:lcvsmstar}
\end{figure*}

In Figure \ref{fig:lcvsmstar}, we show the light curve parameters $\sigma_{\rm BB}$, $\tau_{\rm BB}$, peak bolometric blackbody luminosity $L_{\rm peak, BB}$ calculated using the mean blackbody temperature $T_{\rm 0}$ as a function of the host galaxy stellar mass for the 30 TDEs and AT\,2022cmc obtained from Model 2 (Section \ref{sec:lcfit}). While black hole mass estimates are available for some objects in the comparison sample \citep[e.g.,][]{Hammerstein23_IFU, Yao2023}, we choose the host galaxy stellar mass for consistency across all objects. Additionally, \citet{Hammerstein23_IFU} and \citet{Yao2023} find that the host galaxy stellar mass is strongly correlated with the black hole mass derived from stellar kinematics. We adopt the estimate from \citet{Andreoni2022} for the upper limit on the host stellar mass of $\log(M_{\rm gal}/M_\odot) < 11.2$. We emphasize that this is only an upper limit on the host galaxy mass and may change once better constraints are placed on the host galaxy properties in future observations.

While both the rise and decay timescales for AT\,2022cmc are faster than would normally be expected for a TDE and do not follow the correlation with host mass, this may not be surprising given the light curve and difficulty in constraining a sparsely sampled rise and fast evolving transient. Both the rise and decay timescales are still comparable to other thermal TDEs, albeit for much lower host galaxy masses. The values for the other parameters (peak bolometric luminosity, and blackbody temperature) are more consistent with previously found correlations.

Interestingly, the values for the peak bolometric luminosity and blackbody temperature show remarkable similarity to the TDE-featureless class. \citet{Andreoni2022} noted the potential connection between AT\,2022cmc and the TDE-featureless class due to similarities in photometric and spectroscopic properties, such as peak luminosity and lack of broad lines in the optical spectra. They suggest that the TDE-featureless class may represent the off-axis jetted TDE scenario. This connection is potentially further supported due to the high black hole masses observed for TDE-featureless objects, which could necessitate a non-negligible spin for the black holes in these events so that the stars are disrupted outside of the event horizon, depending on the type of star that is disrupted. \citet{Hammerstein23_IFU} reported a mass of $\log(M_{\rm BH}/M_\odot) = 8.01 \pm 0.82$ obtained from stellar kinematics for AT\,2020qhs. Additional measurements made by \citet{Yao2023} for featureless events yielded masses $\log(M_{\rm BH}/M_\odot)=7.16-8.23$. \citet{Mummery24} found that the black hole masses inferred from the late-time light curve plateaus of TDE-featureless events strongly imply non-negligible spin parameters. If high spin is required to launch a relativistic jet \citep[e.g.,][]{Tchekhovskoy14, Andreoni2022}, then the TDE-featureless class may be capable of launching relativistic jets. This should be interpreted cautiously, however, as TDEs around SMBHs with masses lower than $\sim10^{8} M_\odot$ (the ``Hills mass'') may still have large spins and SMBHs above $\sim10^{8} M_\odot$ can disrupt non-solar type stars outside of the event horizon. 

A high black hole mass for jetted events would be in contrast to the findings of \citet{Eftekhari24}, which disfavored larger black holes for producing jets in TDEs and concluded that jetted TDEs may preferentially come from lower mass SMBHs as a result of lower jet and higher disk efficiencies at higher black hole masses. \citet{Eftekhari24} also found evidence for the cessation of jet activity in the late-time X-ray light curve of AT\,2022cmc, which occurred at $t_{\rm rest} \approx 98$ days. In the optical light curve, we see an unusual steepening in the decay rate approximately 40 days prior to the X-ray break. Our broken power-law fits to the $r$- and $g$-band light curves constrain the time between the optical break and the X-ray break to be $\sim 46$ days. \citet{Eftekhari24} attribute the sudden drop in X-ray emission to an accretion state transition which subsequently manifests as a jet shut-off. The break in the optical light curve prior to the cessation of the jet may imply that the origin of the optical emission is in fact related to the accretion flow or reprocessing of emission related to the accretion flow. However, it is difficult to discern whether the peculiar late-time evolution of the optical light curve is related to the intrinsic evolution of the transient, particularly given the reddening of the SED after $\sim50$ days.

Two other jetted TDE candidates exhibited faint optical counterparts: Sw J2058+05 \citep{Cenko2012, pasham15} and Sw J1112-82 \citep{Brown2015, Brown2017}. \citet{pasham15} fit the UV/optical observations of Sw J2058+05, which also showed a featureless optical spectrum, with a single blackbody model, yielding a mean temperature over several epochs of $ T_{\rm BB} = 10^{4.43}$ K. This is comparable to AT\,2022cmc and the TDE-featureless objects. The peak bolometric luminosity implied by the UV/optical observations is also comparable to AT\,2022cmc and the TDE-featureless class at $L_{\rm BB} = 10^{45.03}$ erg s$^{-1}$. We show this limit in Figure \ref{fig:lcvsmstar}. Black hole mass estimates from host galaxy properties for both of these events are significantly lower than estimates for the mass in AT\,2022cmc \citep{Andreoni2022, pasham15, Brown2017}. As more jetted TDEs are discovered and their host galaxies are characterized, we can further explore the similarities and differences between the optical components in jetted TDEs and the TDE-featureless class.

There have yet to be published observations of TDE-featureless objects that may indicate the presence of a jet, such as late-time radio emission. \citet{Cendes2023} found that $\sim40\%$ of optical TDEs are detected at late times in the radio, but concluded that this emission is likely due to outflows instead of off-axis relativistic jets. It is clear that future and long-term follow-up observations of TDE-featureless events are needed to further explore this possible connection. Interestingly, the rate of ``over-luminous'' TDEs (i.e., $M_r < -20.5$ mag, which includes both jetted TDEs and featureless TDEs) is $2~\rm{Gpc}^{-1}~\rm{yr}^{-1}$ \citep[integrated from the luminosity function in][]{Yao2023}, which is roughly consistent with the estimate from \citet{Andreoni2022} that $\sim1\%$ of all TDEs produce relativistic jets.

Additionally, spectral sequences of featureless TDEs from at or near peak to post-peak are available for only one event \citep[see Figure 7 of][]{Yao2023}, as are spectra outside of the rest-frame optical band. \citet{Guillochon2014} showed that for PS1-10jh, the lines of H$\alpha$, H$\beta$, and \ion{He}{1} were not visible at early times because of the intensity of the hydrogen-ionizing flux, but may have become visible at later times as the event faded. Due to their high luminosities, this may be a plausible explanation for the featureless nature of jetted and TDE-featureless spectra as well as for their high temperatures. Therefore, it may not be required that TDE-featureless events are also jetted but are connected to jetted TDEs by luminosity alone. This could also explain the higher luminosities of some TDE-He events \citep[e.g.,][]{Hammerstein23}. Host galaxy characterization of jetted TDEs may also help to confirm a link to featureless thermal TDEs, whose hosts are typically more massive and redder than other types of thermal TDEs \citep{Hammerstein23, Yao2023}.

We have shown here that the thermal component present in the UV/optical light curve of AT\,2022cmc is indeed comparable to the light curves of other TDEs which are dominated by almost exclusively thermal emission. Moreover, we have shown that the properties of the thermal component are similar to the class of featureless TDEs put forth by \citet{Hammerstein23}. Future studies (e.g., Y.~Yao et al. in prep), will need to investigate the presence of emission that might indicate that these TDE-featureless objects are indeed off-axis analogs to jetted TDEs.

\begin{acknowledgements}
Portions of this work served as a chapter in EH's dissertation. This material is based upon work supported by NASA under award number 80GSFC21M0002. The TReX team at UC Berkeley is partially funded by the National Science Foundation under award No.~AST-2224255. A.R. acknowledge INAF project Supporto Arizona \& Italia.

Based on observations made with the Nordic Optical Telescope, owned in collaboration by the University of Turku and Aarhus University, and operated jointly by Aarhus University, the University of Turku and the University of Oslo, representing Denmark, Finland and Norway, the University of Iceland and Stockholm University at the Observatorio del Roque de los Muchachos, La Palma, Spain, of the Instituto de Astrofisica de Canarias. The GROWTH India Telescope (GIT) is a 70-cm telescope with a 0.7-degree field of view, set up by the Indian Institute of Astrophysics (IIA) and the Indian Institute of Technology Bombay (IITB) with funding from Indo-US Science and Technology Forum and the Science and Engineering Research Board, Department of Science and Technology, Government of India. It is located at the Indian Astronomical Observatory (IAO, Hanle). We acknowledge funding by the IITB alumni batch of 1994, which partially supports the operation of the telescope. The LBT is an international collaboration of the University of Arizona, Italy (INAF: Istituto Nazionale di Astrofisica), Germany (LBTB: LBT Beteiligungsgesellschaft), The Ohio State University, representing also the University of Minnesota, the University of Virginia, and the University of Notre Dame.

\end{acknowledgements}

\bibliography{main, bibliography}{}
\bibliographystyle{aasjournalv7}

\appendix

\section{AT\,2022cmc Light Curve}
In Table \ref{tab:newobs}, we present the new observations of AT\,2022cmc since its discovery in \citet{Andreoni2022}. We note that all observations use the SDSS photometric system, which exhibits slight differences from the ZTF photometric system. This is accounted for when fitting the light curve and SED.

\startlongtable
\begin{deluxetable}{ccccc}
\tablecaption{New observations of AT\,2022cmc.}
\tablehead{\colhead{MJD} & \colhead{Filter} & \colhead{Mag} & \colhead{eMag} & \colhead{Instrument}}
\startdata
59669.87 & 	\textit{r} & 22.04 & 0.11 & GITCamera \\
59676.79 & 	\textit{r} & 22.21 & 0.1 & GITCamera \\
59676.87 & 	\textit{g} & 22.02 & 0.11 & GITCamera \\
59677.1 & 	\textit{g} & 22.11 & 0.06 & ALFOSC \\
59677.11 & 	\textit{r} & 22.12 & 0.06 & ALFOSC \\
59677.13 & 	\textit{i} & 22.22 & 0.07 & ALFOSC \\
59689.98 & 	\textit{r} & 22.28 & 0.16 & IO:O \\
59699.95 & 	\textit{g} & 22.51 & 0.17 & IO:O \\
59703.96 & 	\textit{g} & 22.58 & 0.07 & ALFOSC \\
59703.97 & 	\textit{r} & 22.5 & 0.08 & ALFOSC \\
59703.99 & 	\textit{i} & 22.53 & 0.1 & ALFOSC \\
59721.99 & 	\textit{g} & 23.09 & 0.1 & ALFOSC \\
59722.0 & 	\textit{r} & 23.03 & 0.11 & ALFOSC \\
59722.02 & 	\textit{i} & 22.94 & 0.12 & ALFOSC \\
59732.22 & 	\textit{g} & 23.18 & 0.06 & LMI \\
59732.22 & 	\textit{r} & 23.06 & 0.06 & LMI \\
59732.22 & 	\textit{i} & 22.88 & 0.06 & LMI \\
59752.98 & 	\textit{g} & 23.82 & 0.14 & ALFOSC\\
59753.0 & 	\textit{r} & 23.48 & 0.12 & ALFOSC \\
59753.01 & 	\textit{i} & 23.46 & 0.14 & ALFOSC \\
59767.91 & 	\textit{r} & 25.8 & 0.58 & ALFOSC \\
59767.91 & 	\textit{g} & 26.77 & 99.0 & ALFOSC \\
59767.92 & 	\textit{i} & 25.41 & 0.48 & ALFOSC \\
59769.0 & 	\textit{i} & 23.2 & 0.2 & LMI \\
59778.9 & 	\textit{r} & 25.05 & 0.26 & ALFOSC \\
59780.93 & 	\textit{i} & 24.44 & 0.26 & ALFOSC \\
59674.0 & 	\textit{u} & 22.14 & 0.02 & LBC \\
59674.0 & 	\textit{g} & 22.15 & 0.02 & LBC \\
59674.0 & 	\textit{r} & 22.08 & 0.03 & LBC \\
59674.0 & 	\textit{i} & 22.1 & 0.04 & LBC \\
59674.0 & 	\textit{z} & 22.2 & 0.07 & LBC \\
59690.0 & 	\textit{u} & 22.59 & 0.07 & LBC \\
59690.0 & 	\textit{g} & 22.36 & 0.06 & LBC \\
59690.0 & 	\textit{r} & 22.36 & 0.06 & LBC \\
59690.0 & 	\textit{i} & 22.35 & 0.08 & LBC \\
59690.0 & 	\textit{z} & 22.51 & 0.15 & LBC \\
59729.0 & 	\textit{u} & 23.36 & 0.05 & LBC \\
59729.0 & 	\textit{g} & 23.16 & 0.04 & LBC \\
59729.0 & 	\textit{r} & 22.94 & 0.04 & LBC \\
59729.0 & 	\textit{i} & 22.94 & 0.08 & LBC \\
59729.0 & 	\textit{z} & 22.82 & 0.16 & LBC \\
59768.0 & 	\textit{u} & 24.71 & 0.35 & LBC \\
59768.0 & 	\textit{g} & 24.2 & 0.19 & LBC \\
59768.0 & 	\textit{r} & 23.79 & 0.15 & LBC \\
59768.0 & 	\textit{i} & 23.61 & 0.16 & LBC \\
59768.0 & 	\textit{z} & 23.21 & 0.21 & LBC \\
59732.2 & 	\textit{g} & 23.18 & 0.06 & LMI \\
59732.21 & 	\textit{r} & 23.06 & 0.06 & LMI \\
59732.22 & 	\textit{i} & 22.88 & 0.06 & LMI \\
59769.23 & 	\textit{i} & 23.19 & 0.14 & LMI \\
\enddata
\tablecomments{New observations of AT2022cmc since its discovery in \citet{Andreoni2022}.}
\label{tab:newobs}
\end{deluxetable}

\end{document}